%% file: ChicjTo2Lambda2Pi.tex
\documentclass[prd,amsmath,amssymb,showpacs,superscriptaddress,nofootinbib, 12pt]{revtex4-1}
\usepackage{amssymb}
\usepackage{graphicx,epsfig}
\usepackage{dcolumn}
\usepackage{bm}
\usepackage[dvipdfm,CJKbookmarks=true,unicode,colorlinks,linkcolor=blue,anchorcolor=blue,citecolor=blue,pdfborder={0 0 0}]{hyperref}
\usepackage{amsfonts}
\usepackage{keyval,graphicx}
\usepackage{textcomp,wasysym}

\usepackage[english]{babel}
\usepackage{overpic}
\usepackage{multirow}
\usepackage{supertabular}
\usepackage{amsmath}
\usepackage{longtable}
\usepackage{enumerate}

\usepackage[flushleft]{caption2}

\newcommand{\eg}{\emph{e}.\emph{g}.}
\newcommand{\ie}{\emph{i}.\emph{e}.}

\usepackage{array}
\newcommand{\PreserveBackslash}[1]{\let\temp=\\#1\let\\=\temp}
\newcolumntype{C}[1]{>{\PreserveBackslash\centering}p{#1}}
\newcolumntype{R}[1]{>{\PreserveBackslash\raggedleft}p{#1}}
\newcolumntype{L}[1]{>{\PreserveBackslash\raggedright}p{#1}}

\begin{document}

\normalsize
\parskip=5pt plus 1pt minus 1pt

\title{\boldmath Observation of $\chi_{cJ}$ Decays to $\Lambda\bar{\Lambda}\pi^{+}\pi^{-}$}

\author{
  {\small
    M.~Ablikim$^{1}$, M.~N.~Achasov$^{5}$, D.~J.~Ambrose$^{39}$, F.~F.~An$^{1}$, Q.~An$^{40}$,
      Z.~H.~An$^{1}$, J.~Z.~Bai$^{1}$, Y.~Ban$^{27}$, J.~Becker$^{2}$, J.~V.~Bennett$^{17}$
% ...
	\\
	\vspace{0.2cm}
    (BESIII Collaboration)\\
      \vspace{0.2cm} {\it
	$^{1}$ Institute of High Energy Physics, Beijing 100049, People's Republic of China\\
	  $^{2}$ Bochum Ruhr-University, 44780 Bochum, Germany\\
% ...
$^{5}$ G.I. Budker Institute of Nuclear Physics SB RAS (BINP), Novosibirsk 630090, Russia\\
% ...
$^{17}$ Indiana University, Bloomington, Indiana 47405, USA\\
% ...
$^{27}$ Peking University, Beijing 100871, People's Republic of China\\
% ...
$^{39}$ University of Rochester, Rochester, New York 14627, USA\\
	  $^{40}$ University of Science and Technology of China, Hefei 230026, People's Republic of China\\
% ...
      }
  }
  \vspace{0.4cm}
}
\date{\today}

\begin{abstract}
Decays of the $\chi_{cJ}$ states ($J$=0,\ 1,\ 2) to
$\Lambda\bar{\Lambda}\pi^{+}\pi^{-}$,  including
processes with intermediate $\Sigma(1385)$, are studied
through the $E$1 transition $\psi^\prime\to\gamma\chi_{cJ}$ using 106
million $\psi^\prime$ events collected with the BESIII detector at
BEPCII.  This is the first observation of $\chi_{cJ}$ decays to the
final state $\Lambda\bar{\Lambda} \pi^{+}\pi^{-}$. The branching ratio
of the intermediate process $\chi_{cJ}\rightarrow\Sigma(1385)^{\pm}
\bar{\Sigma}(1385)^{\mp}$ is also measured for the first time, and
the results  agree with the theoretical predictions based on the
color-octet effect.
\end{abstract}


\maketitle

\section{Introduction}
Decays of $P$-wave charmonium states, \eg, the $\chi_{cJ}$, cannot be well
explained  by the color-singlet contribution alone, although this works
well in explaining the decays of $S$-wave charmonium,  \eg, the J/$\psi$
and $\psi^\prime$.  In calculations of the color-octet contribution, %
Ref.~\cite{PRAv674p185} predicted branching ratios of
$\chi_{cJ}\to$baryon$+$anti-baryon in which  the $\chi_{cJ}\rightarrow
p \bar{p}$ result is consistent with experimental observation, while the
$\chi_{cJ}\rightarrow\Lambda\bar{\Lambda}$~\cite{pdg12} result is not.
The calculated branching ratios are
$\mathcal{B}(\chi_{c1}\rightarrow \Lambda\bar{\Lambda}) = (3.91\pm0.24)\times10^{-5}$ and
$\mathcal{B}(\chi_{c2} \rightarrow\Lambda\bar{\Lambda}) = (3.49\pm0.20)\times10^{-5}$, %
while the experimental results are $(11.8\pm1.9)\times10^{-5}$ and
$(18.6\pm2.7) \times10^{-5}$, respectively.
In addition to $\Lambda\bar{\Lambda}$, reference~\cite{PRAv674p185} also calculated
the branching ratios of $\chi_{c1}\rightarrow\Sigma(1385)\bar{\Sigma}(1385)$
and $\chi_{c2}\rightarrow\Sigma(1385)\bar{\Sigma}(1385)$ to be
$(2.15\pm0.12)\times10^{-5}$ and $(3.61\pm0.20)\times10^{-5}$, %
respectively, but there are no previous experimental results on these decay channels.
Therefore, it is meaningful to test these predictions experimentally.
In addition, due to the helicity selection rule, the decay of $\chi_{c0}$
into baryon-antibaryon is expected to be suppressed~\cite{hsl}.

In this paper, we report measurements of
$\chi_{cJ}\rightarrow\Lambda\bar{\Lambda}\pi^{+}\pi^{-}$
($J=0,\ 1,\ 2$) (including the intermediate $\Sigma(1385)$ resonance), %
$\chi_{cJ}\rightarrow\Sigma(1385)^{\pm}\bar{\Lambda}\pi^{\mp}+c.c.$, %
and  $\chi_{cJ}\rightarrow\Sigma(1385)^{\pm}  \bar{\Sigma}(1385)^{\mp}$
through the $E$1 transition  $\psi^\prime\rightarrow\gamma\chi_{cJ}$, %
where $\Sigma(1385)^{\pm}\rightarrow\Lambda\pi^{\pm}$ and
$\Lambda\rightarrow p\pi^{-}$.  This  work is based on a $106$
million $\psi^{\prime}$ event sample collected  with the BESIII detector
at the Beijing Electron-Positron Collider II (BEPCII)~\cite{BAM0003}.
Continuum data taken  at the center of mass energy $\sqrt{s}=3.65$\,GeV, with
an integrated luminosity of $42.9$\,$\mathrm{pb}^{-1}$,  is used to study
non-$\psi^{\prime}$ decay background.

\section{The BESIII Detector}
BEPCII~\cite{bes3} is a double-ring, multi-bunch $e^+e^-$ collider with
collision energies ranging from 2.0\,GeV to 4.6\,GeV. The BESIII
detector~\cite{bes3} is a general-purpose spectrometer with 93\%
coverage of full solid angle. From the interaction point outwards,
	 BESIII is composed of the following: a main drift chamber consisting
	 of 43 layers of drift cells with a space resolution of about
	 135\,$\mu$m and momentum resolution of about $0.5\%$ at 1\,GeV/c; a
	 time-of-flight counter, which is comprised of two layers of
	 scintillator with time resolution of 80\,ps in the barrel part and
  one layer with time resolution of 110\,ps in the end-cap part; an
electromagnetic calorimeter (EMC), which is comprised of 6240 CsI(Tl)
  crystals, with energy resolution of $2.5\%$ in the barrel and $5.0\%$
  in the end-cap for a 1\,GeV photon, and position resolution of 6\,mm
  in the barrel and 9\,mm in the end-cap; a super-conducting solenoid
  magnet, which can provide a 1\,T magnetic field parallel to the beam
  direction; and a muon counter, which is made of 1000\,m$^{2}$
  resistive-plate-chambers sandwiched in iron absorbers.

  \section{Monte-Carlo Simulation}
  For evaluation of the detection efficiency and understanding
  backgrounds, a Monte-Carlo (MC) simulation framework for BESIII was
  developed. A GEANT4-based MC simulation program, BOOST, is
  designed to simulate the interaction of particles in the spectrometer
  and the responses of the detector. For the generation of charmonium
  states, \eg, $\psi^\prime$, an event generator,  %
  KKMC~\cite{kkmc_1, kkmc_2}, is employed, which handles the initial
  state radiative  correction and the beam energy spread. For
  simulation of the resonant decay, BesEvtGen, based on
  EvtGen~\cite{besevtgen_1, besevtgen_2}, is used to realize well-measured
  processes, while LundCharm~\cite{besevtgen_1} is used for the unknown
  possible processes.

  In the MC simulations for the processes presented here, %
  $\psi^\prime\rightarrow\gamma\chi_{cJ}$ is assumed to be a pure $E$1
  transition, and the  polar angle, $\theta$, follows a distribution
  of the form $1+\alpha\cos^{2}\theta$ with $\alpha$ $=$ 1, $-1/3$, and
  $1/13$ for $J$ $=$ 0, 1 and 2, respectively~\cite{chicj_decay}.
  Momenta in the decay of
  $\chi_{cJ}\rightarrow\Sigma(1385)^{\pm}\bar{\Sigma}(1385)^{\mp}$, %
  $\chi_{cJ}\rightarrow\Sigma(1385)^{\pm}\bar{\Lambda}\pi^{\mp}(c.c.)$ and
  $\chi_{cJ}\rightarrow\Lambda\bar{\Lambda}\pi^{+}\pi^{-}$
  are uniformly distributed in phase space. For the decay mode
  $\chi_{cJ}\rightarrow \Sigma(1385)^{\pm}\bar{\Sigma}(1385)^{\mp}$, an
  extreme angular distribution is used to test the phase space assumption
  and no significant differences in efficiencies are observed. This is because, with
   the current level of statistics, the detection efficiencies of the final states are determined mainly by the detection of the $E1$ photons, and the angular distributions of the hadrons in the subsequent decays are not dependent on their MC decay models.

  \section{Event Selection}\label{event_selection}

  The candidate events for the decay modes $\psi^\prime\rightarrow\gamma
  \chi_{cJ} \rightarrow\gamma\Lambda\bar{\Lambda}\pi^{+}\pi^{-}$, with
  $\Lambda\to p\pi$, were chosen with the following selection criteria:

  (1) Charged tracks, \ie, candidates for $\pi^{\pm}$, $p$ and
  $\bar{p}$, must satisfy $\left| \cos\theta \right| \le 0.93$, where
  $\theta$ is the polar angle with respect to the beam direction.
  Particle identification is not used.

  (2) The charged tracks not assigned to any $\Lambda(\bar{\Lambda})$
  decay candidates must have their point of closest approach to the
  interaction point within 10\,cm along the beam direction and 1\,cm
  in the perpendicular plane.

  (3) A common vertex constraint is applied to each pair of charged tracks
  assumed to decay from $\Lambda/\bar{\Lambda}$  \ie, $p\pi^{-}$ and
  $\bar{p}\pi^{+}$, and the production points of  $\Lambda/\bar{\Lambda}$
  candidates are constrained to the interaction point.

  (4) A photon candidate is a shower cluster in the EMC that is not
  associated with any charged track and has a  minimum energy deposit
  of 25\,MeV in the barrel or 50\,MeV in the end-cap.

  (5) The total momentum of all final particle candidates is
  constrained to the initial four-momentum of the $e^{+}e^{-}$ system in
  a kinematic fit.  The events with $\chi^{2}_{4\mathrm{C}}<80$ are
  retained; for an event with more than one photon candidate, only the
  one with the smallest $\chi^{2}_{4\mathrm{C}}$ is kept.

  (6) Backgrounds from the decay $\psi^\prime\rightarrow\pi^{+}\pi^{-}\mathrm{J}/\psi$
  followed by J/$\psi\rightarrow\gamma\Lambda\bar{\Lambda}$ are rejected by requiring
  the  $\pi^{+}\pi^{-}$ recoil mass be greater than  3.108\,GeV/$c^{2}$ or less
  than 3.088\,GeV/$c^{2}$. The background from
  $\psi^\prime\rightarrow\Sigma(1385)\bar{\Sigma}(1385)$, followed by
  $\Sigma(1385)\rightarrow\Sigma^{0}\pi$ and $\Sigma^{0}\rightarrow\gamma\Lambda$, is
  rejected by discarding events with $\gamma\Lambda$ ($\gamma\bar{\Lambda}$) mass
  in the range [1.183, 1.202]\,GeV/$c^{2}$.

  \section{Signal Estimation}\label{data_analysis}
\begin{figure}[t!]
\begin{minipage}[t]{0.45\linewidth}
\begin{overpic} [height=0.20\textheight,width=1.0\linewidth]
{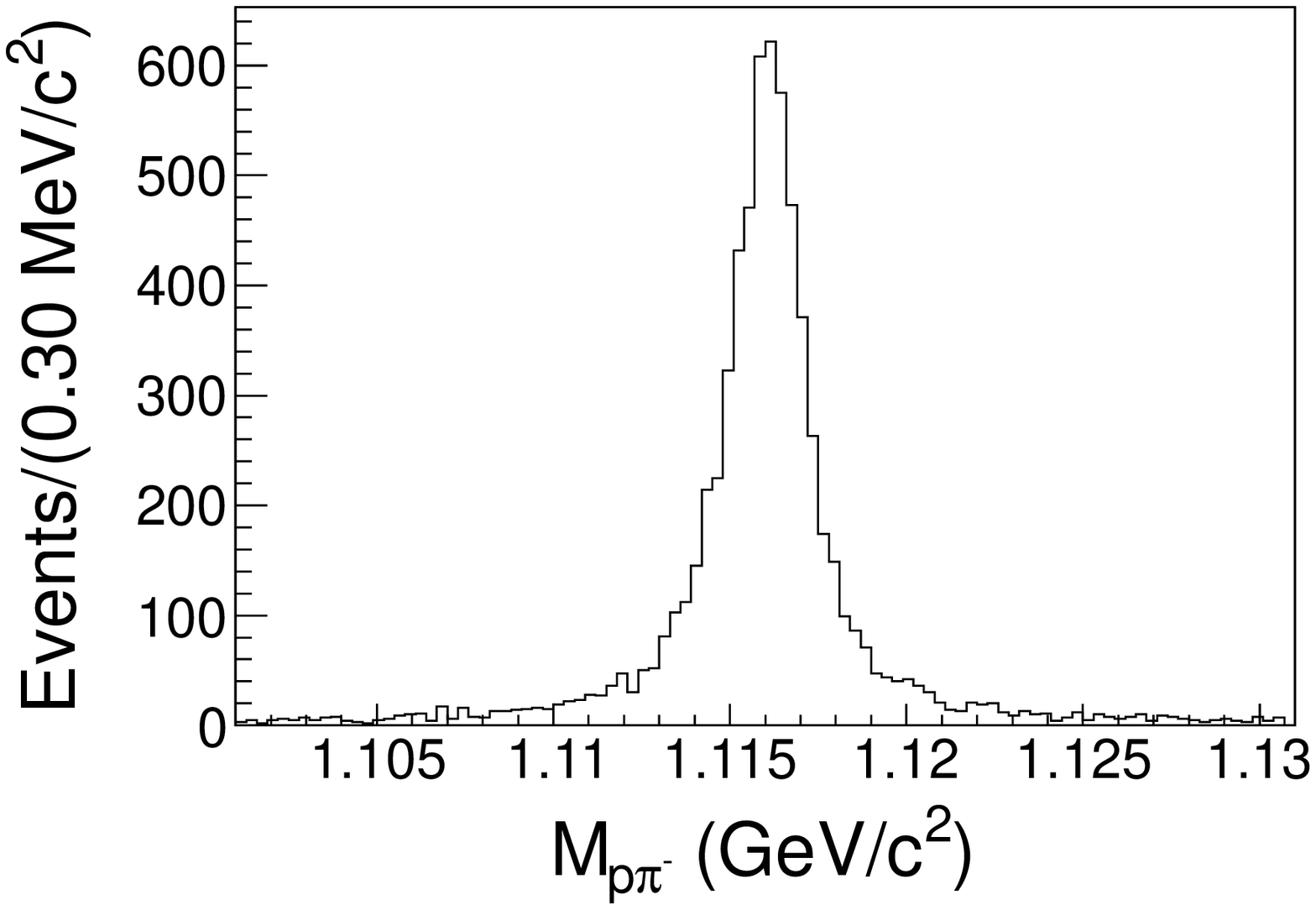}
\put(25, 40){(a)}
\end{overpic}
\end{minipage}
\begin{minipage}[t]{0.45\linewidth}
\begin{overpic} [height=0.20\textheight,width=1.0\linewidth]
{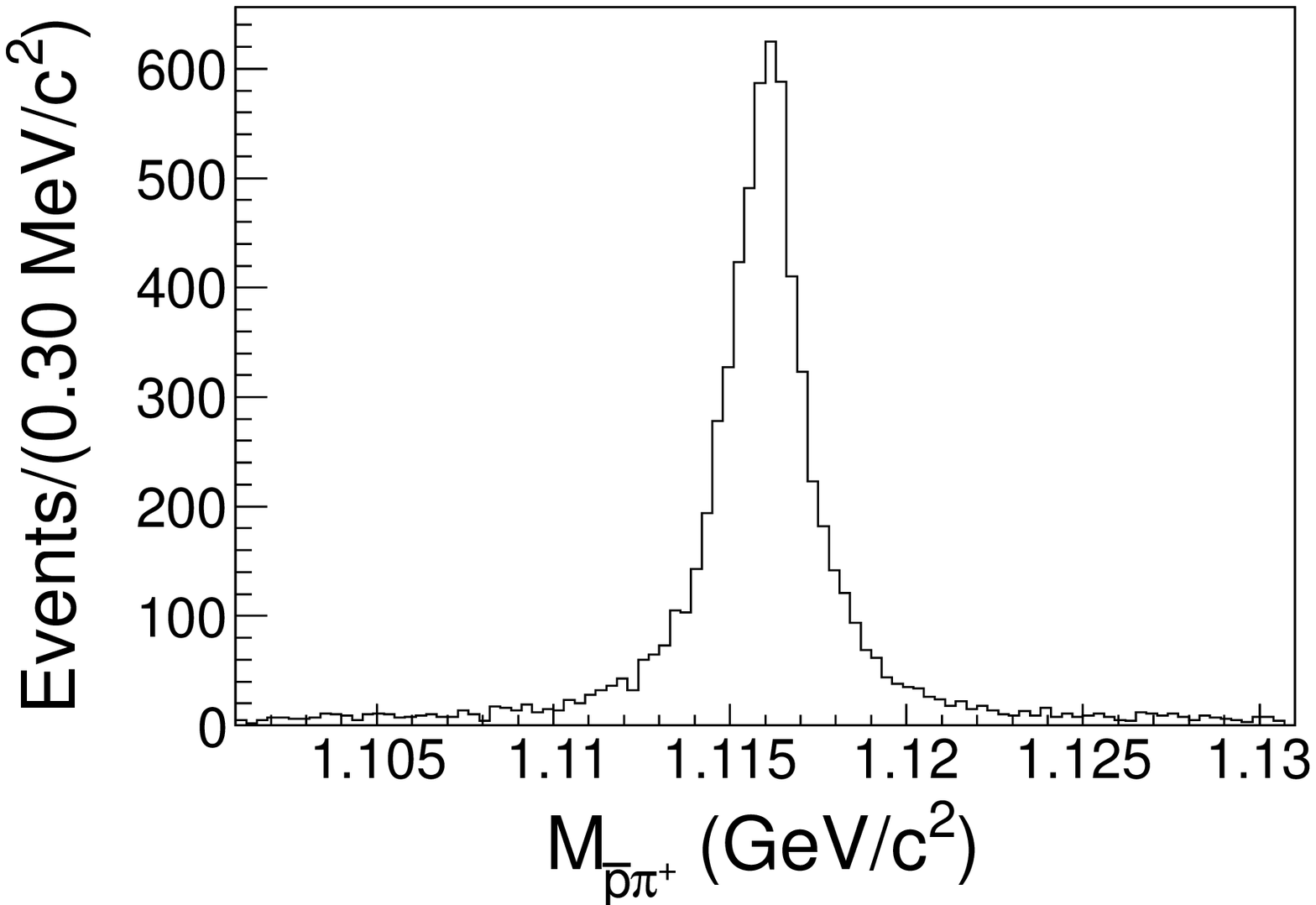}
\put(25, 40){(b)}
\end{overpic}
\end{minipage}
\begin{minipage}[t]{0.45\linewidth}
\begin{overpic} [height=0.20\textheight,width=1.0\linewidth]
{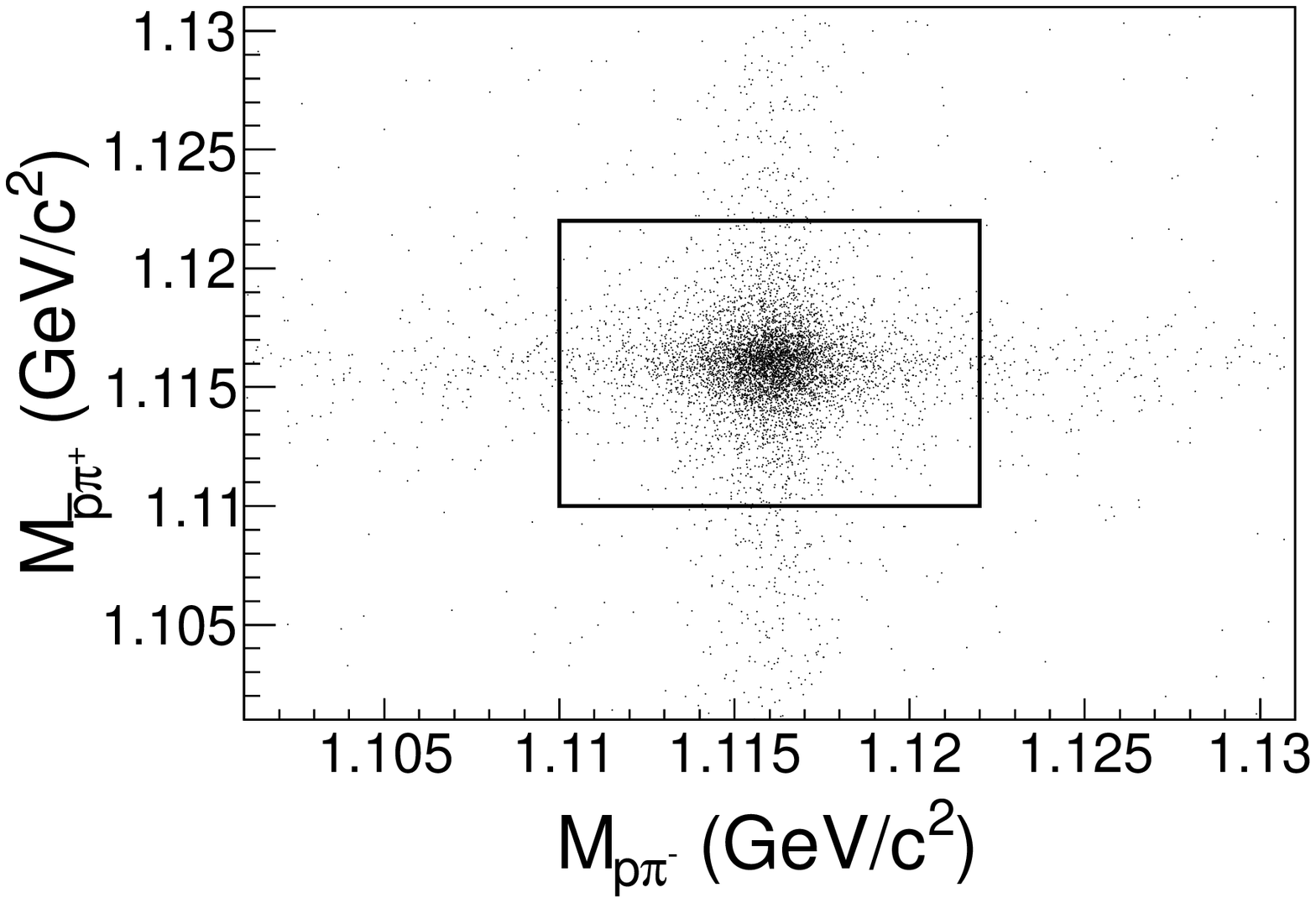}
\put(25, 40){(c)}
\end{overpic}
\end{minipage}
\begin{minipage}[t]{0.45\linewidth}
\begin{overpic} [height=0.20\textheight,width=1.0\linewidth]
{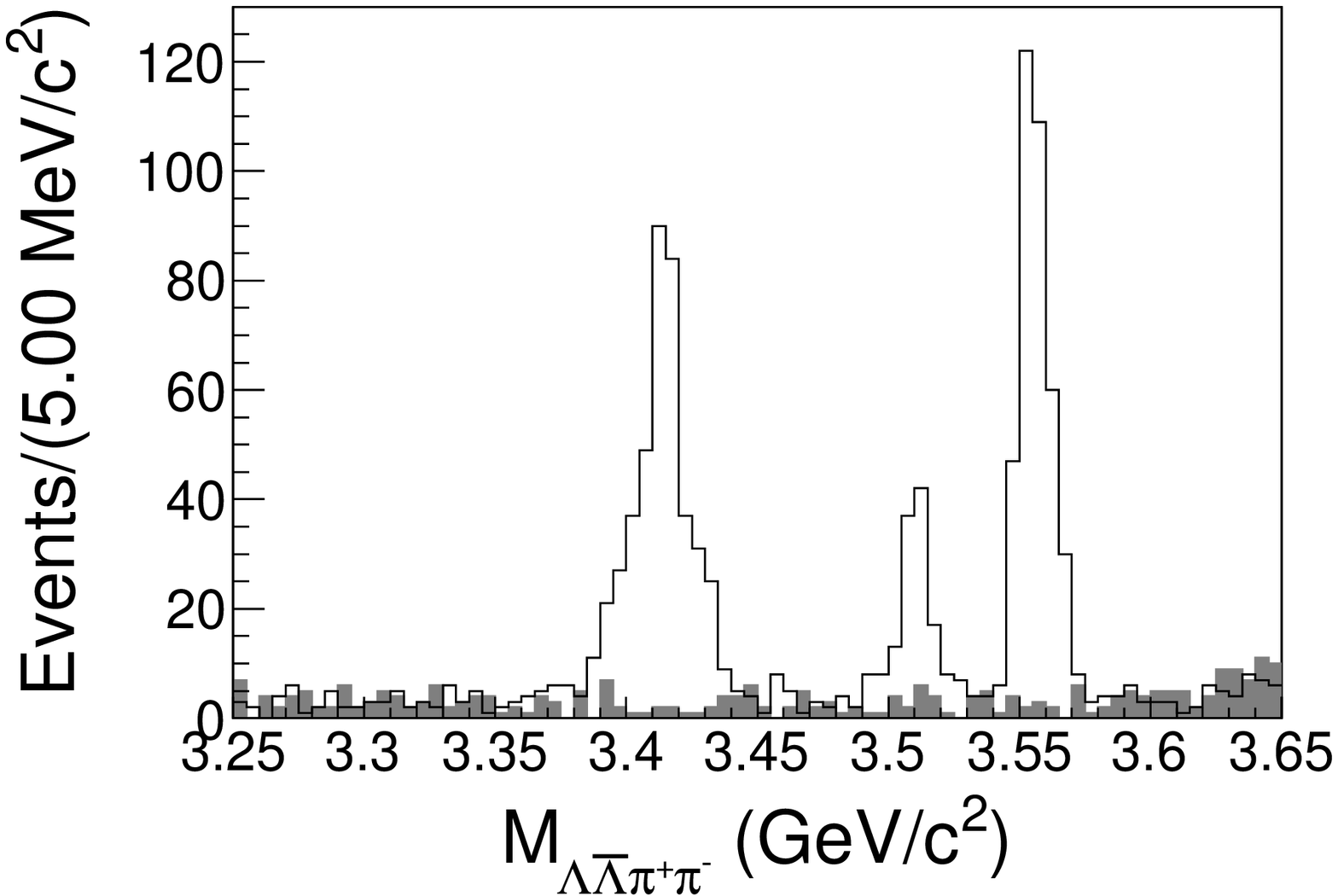}
\put(25, 40){(d)}
\end{overpic}
\end{minipage}
\renewcommand{\figurename}{FIG.}
\caption{
  (a) The invariant mass distribution {\it{M}}$_{p\pi^{-}}$ for $p\pi^{-}$.
    (b) The invariant mass distribution {\it{M}}$_{\bar{p}\pi^{+}}$ for
    $\bar{p}\pi^{+}$. (c) The scatter plot of {\it{M}}$_{p\pi^{-}}$
    versus {\it{M}}$_{\bar{p}\pi^{+}}$; the box indicates the
    $\Lambda\bar{\Lambda}$ signal region used in this analysis.
    (d) The invariant mass distribution {\it{M}}$_{\Lambda\bar{\Lambda}\pi^{+}\pi^{-}}$
    for $\Lambda\bar{\Lambda}\pi^{+}\pi^{-}$; the shaded histogram is the
      background estimated from the inclusive decays of the $\psi^{\prime}$ MC sample.
} {\label{figure_signal_lam_chicj}}
\end{figure}

  The invariant mass distributions of {\it{M}}$_{p\pi^{-}}$ and
{\it{M}}$_{\bar{p}\pi^{+}}$ are shown in Figs.~\ref{figure_signal_lam_chicj}(a) and
(b), where the signals of $\Lambda$ and $\bar{\Lambda}$ are  clean.
Figure~\ref{figure_signal_lam_chicj}(c) shows a scatter
plot ($M_{p\pi^{-}}$ $versus$ $M_{\bar{p}\pi^{+}}$).
Events where {\it{M}}$_{p\pi^{-}}$ and {\it{M}}$_{\bar{p}\pi^{+}}$ fall within
the box in Fig.~\ref{figure_signal_lam_chicj}(c) are used for further analysis.
The invariant mass distribution of $\Lambda\bar{\Lambda}\pi^{+}\pi^{-}$, %
{\it{M}}$_{\Lambda\bar{\Lambda}\pi^{+}\pi^{-}}$, is shown in Fig.
\ref{figure_signal_lam_chicj}(d), and the three $\chi_{cJ}$ peaks  are clearly
observed.

\begin{figure}[t!]
  \begin{minipage}[t]{0.45\linewidth}
  \begin{overpic} [height=0.20\textheight,width=1.0\linewidth]
{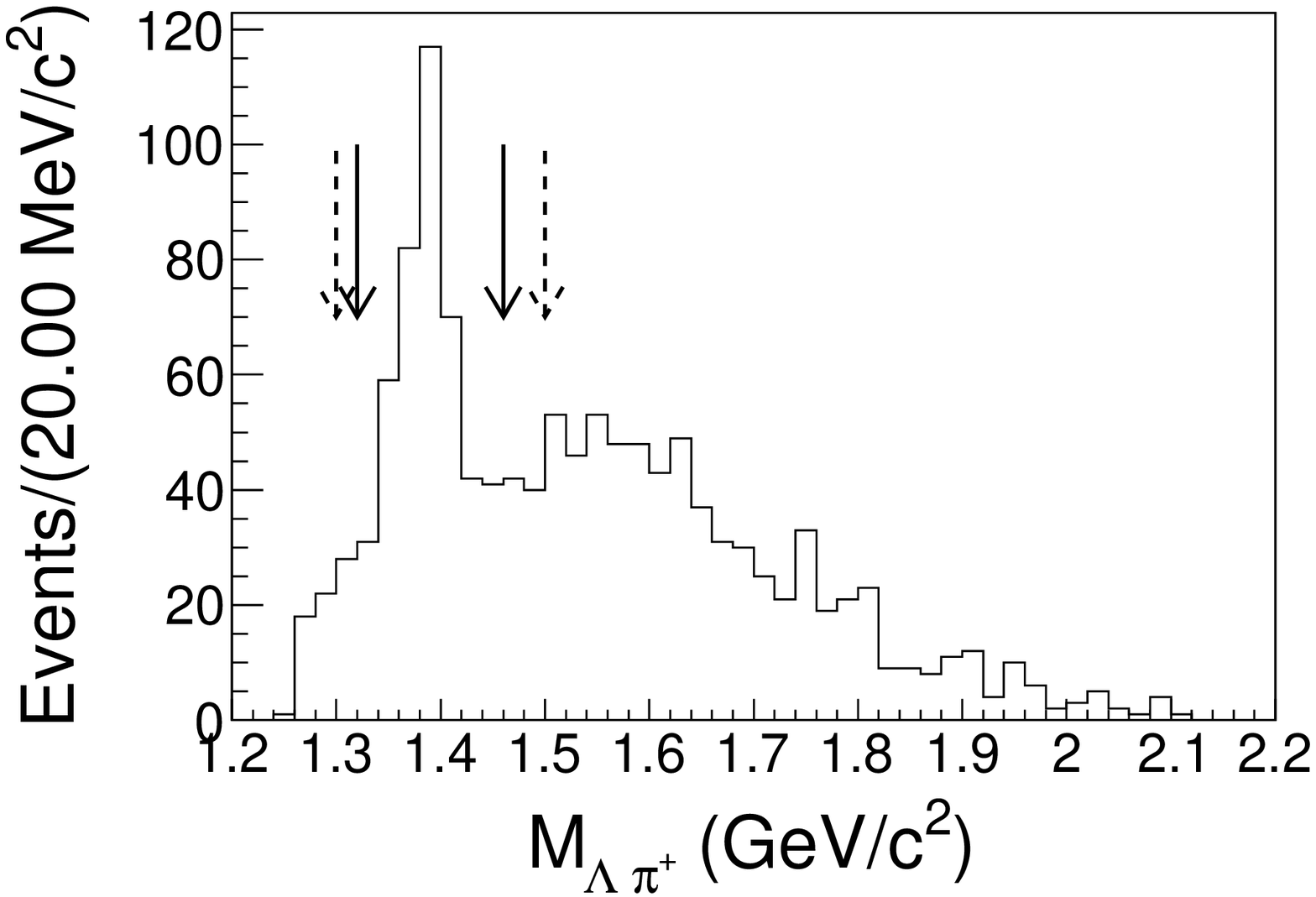}
\put(70, 40){(a)}
\end{overpic}
\end{minipage}
\begin{minipage}[t]{0.45\linewidth}
\begin{overpic} [height=0.20\textheight,width=1.0\linewidth]
{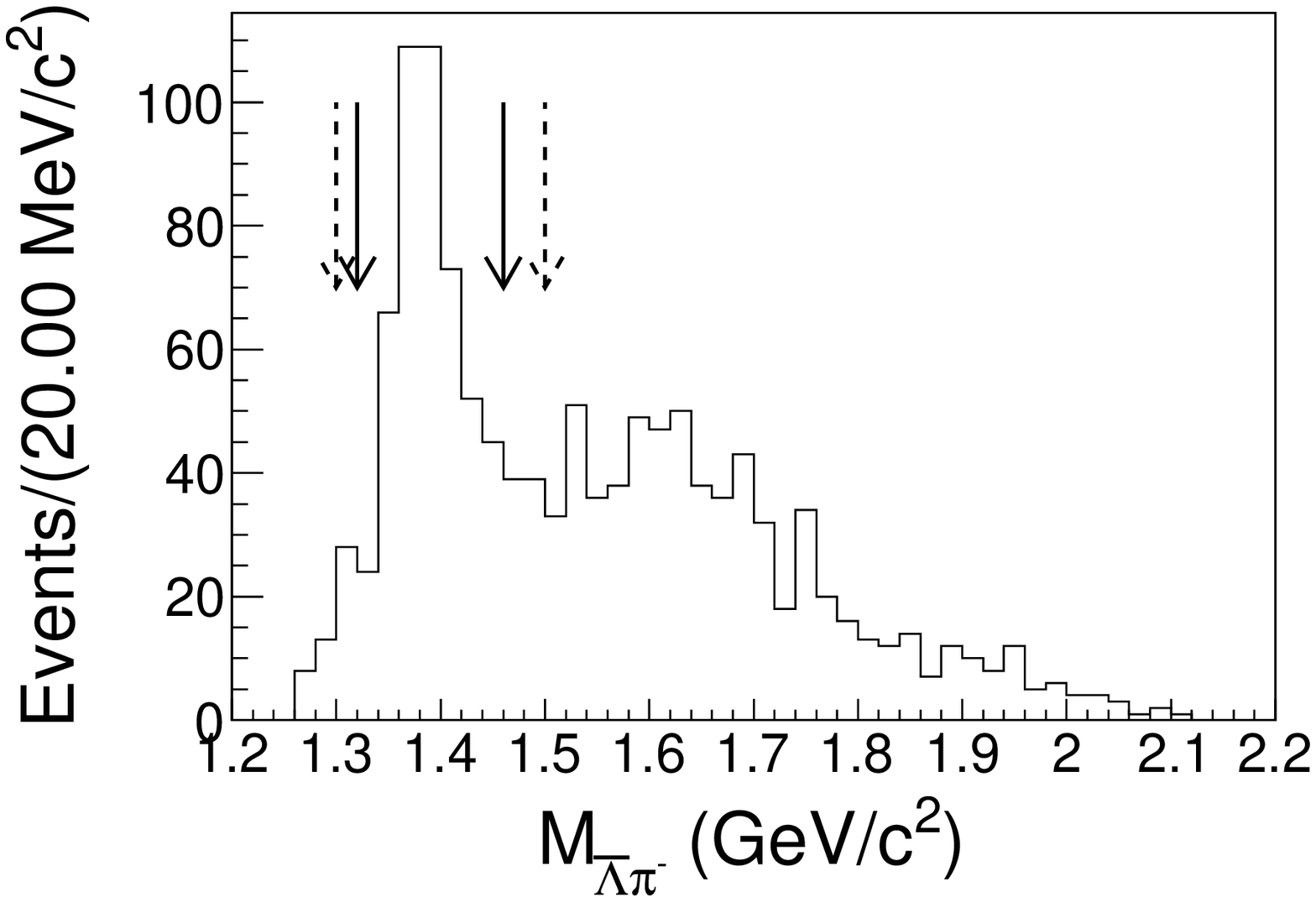}
\put(70, 40){(b)}
\end{overpic}
\end{minipage}
\begin{minipage}[t]{0.45\linewidth}
\begin{overpic} [height=0.20\textheight,width=1.0\linewidth]
{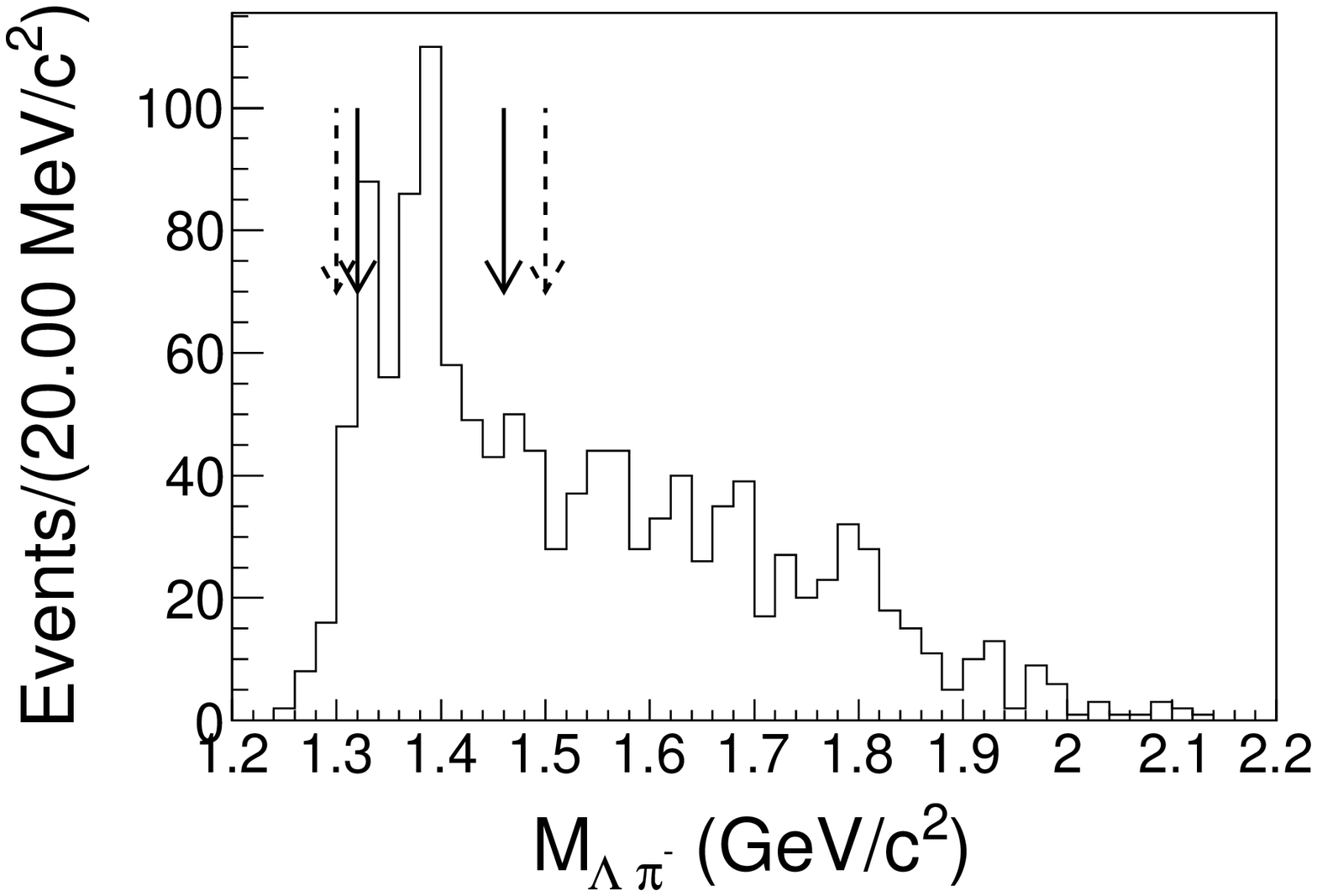}
\put(70, 40){(c)}
\end{overpic}
\end{minipage}
\begin{minipage}[t]{0.45\linewidth}
\begin{overpic} [height=0.20\textheight,width=1.0\linewidth]
{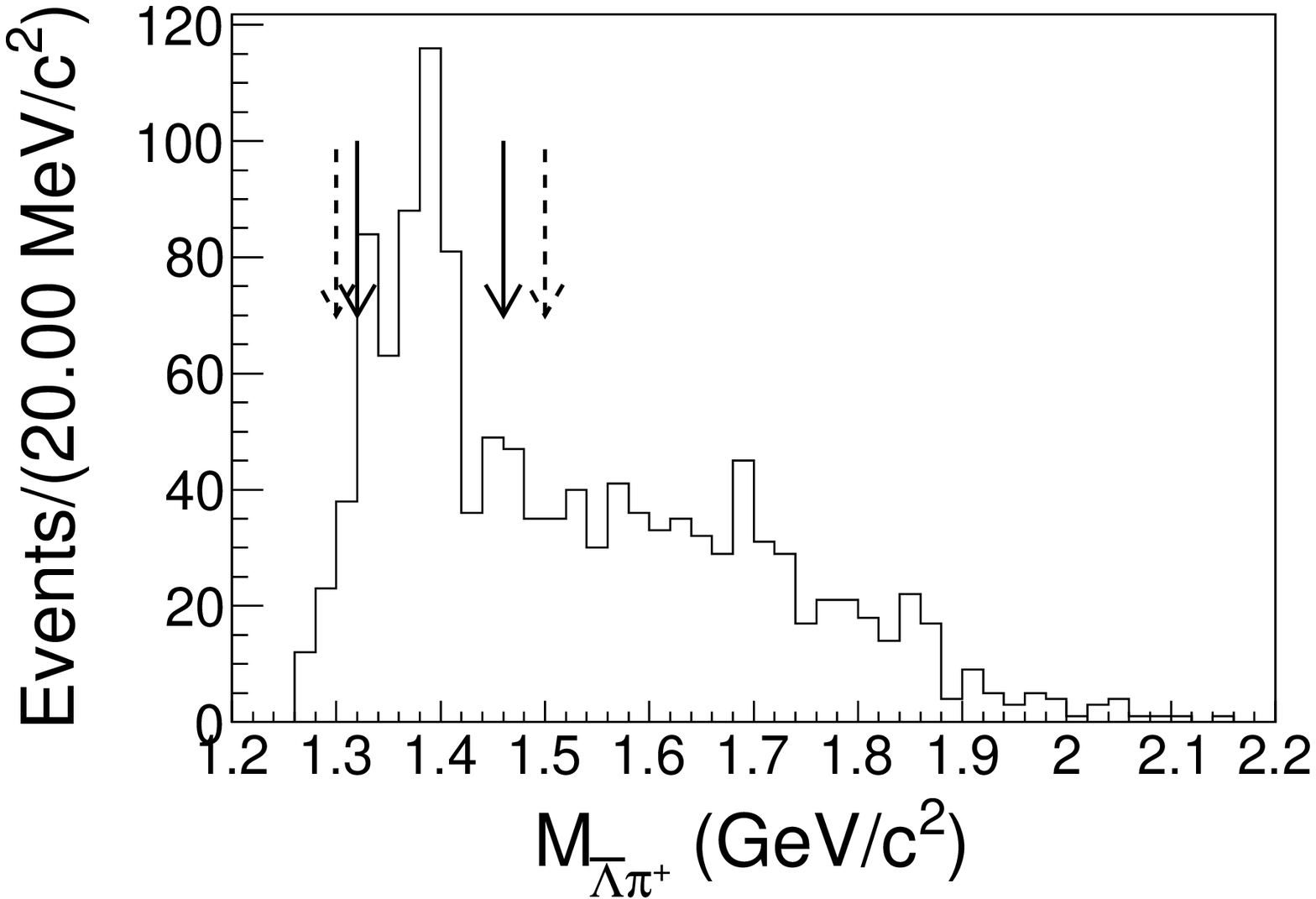}
\put(70, 40){(d)}
\end{overpic}
\end{minipage}
\renewcommand{\figurename}{FIG.}
\caption{
  The  invariant mass distributions of (a) $\Lambda\pi^{+}$, %
    (b) $\bar{\Lambda}\pi^{-}$, (c) $\Lambda\pi^{-}$ and (d)$\bar{\Lambda}\pi^{+}$.
    The areas between the two solid arrows  are taken as the  $\Sigma(1385)$ signal
    regions, while the areas outside the two dashed arrows are non-$\Sigma(1385)$
    regions.  The peaks of $\Xi^{\pm}$ in (c) and (d) will be rejected with the
    requirement $M_{\Lambda\pi^{-}(\bar{\Lambda}\pi^{+})}>$1.331\,GeV/$c^{2}$
    or $M_{\Lambda\pi^{-}(\bar{\Lambda}\pi^{+})}<$1.312\,GeV/$c^{2}$.
}
{\label{figure_signal_s1385}}
\end{figure}

The invariant masses of $\Lambda\pi^{+}$ and $\Lambda\pi^{-}$
  ($\bar{\Lambda}\pi^{-}$ and $\bar{\Lambda}\pi^{+}$) are displayed in
Figs. \ref{figure_signal_s1385}(a) and (c)
  (Figs. \ref{figure_signal_s1385}(b) and (d)), respectively.
  $\Sigma(1385)$ peaks are clearly seen.  $\Xi^{\pm}$ peaks are also
  seen in Figs. \ref{figure_signal_s1385}(c) and (d), %
  around its nominal mass 1.322\,GeV/$c^{2}~$\cite{pdg12}.  The $\Xi$ is
  a relatively long-lived particle, and the selection criteria in this
  analysis are not optimized for a study of the $\Xi$.  Hence, this work
  does not include study of processes involving $\Xi$. Events around the
  $\Xi^{\pm}$ peaks are rejected by requiring
  $M_{\Lambda\pi^{-}(\bar{\Lambda}\pi^{+})}$ be less than
  1.331\,GeV/$c^{2}$ or greater than 1.312\,GeV/$c^{2}$.

  We divide the remaining $\chi_{cJ}$ decays into five processes: (1)~$\Lambda\bar{\Lambda}
  \pi^{+}\pi^{-}$ (non-resonant); (2)~$\Sigma(1385)^{+}\bar{\Lambda}\pi^{-}+c.c.$; %
  (3)~$\Sigma(1385)^{-}\bar{\Lambda}\pi^{+} +c.c.$; %
  (4)~$\Sigma(1385)^{+}\bar{\Sigma} (1385)^{-}$; and
  (5)~$\Sigma(1385)^{-}\bar{\Sigma}(1385)^{+}$. To study the five processes, requirements on
  $M_{\Lambda\pi^{-}(\bar{\Lambda}\pi^{+})}$ are implemented as shown in
  Fig. \ref{figure_signal_s1385}. The areas between 1.32\,GeV/$c^{2}$ and
  1.46\,GeV/$c^{2}$ (two solid arrows) are defined as $\Sigma(1385)$ signal
  regions, while the areas smaller than 1.30\,GeV/$c^{2}$ or larger than
  1.50\,GeV/$c^{2}$ (two dashed arrows) are defined as non-$\Sigma(1385)$
  regions.

\begin{figure}[t!]
\begin{minipage}[t]{0.45\linewidth}
\begin{overpic}[height=0.20\textheight,width=1.0\linewidth]
{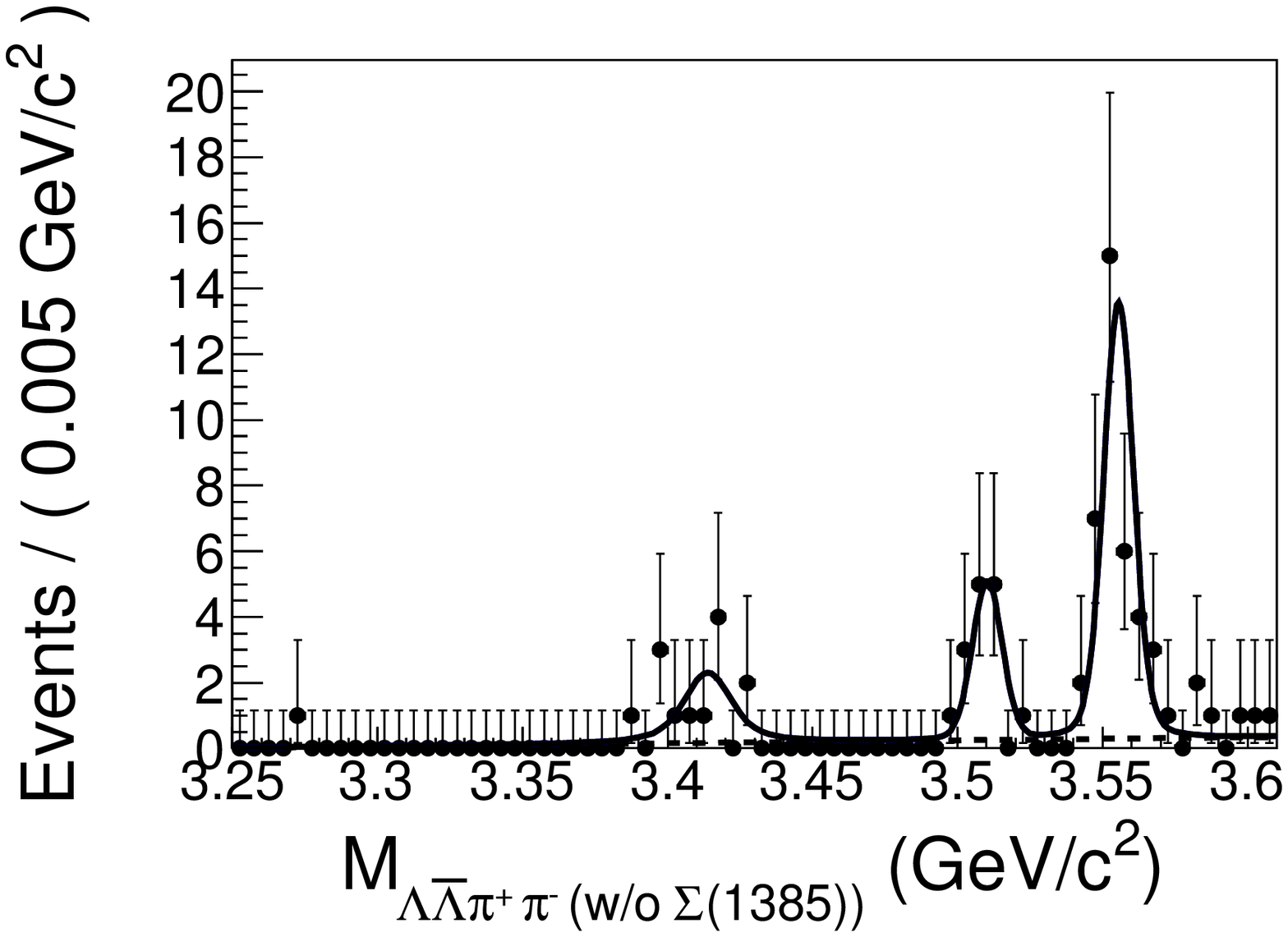}
\put(25, 40){(a)}
\end{overpic}
\end{minipage}
\begin{minipage}[t]{0.45\linewidth}
\begin{overpic}[height=0.20\textheight,width=1.0\linewidth]
{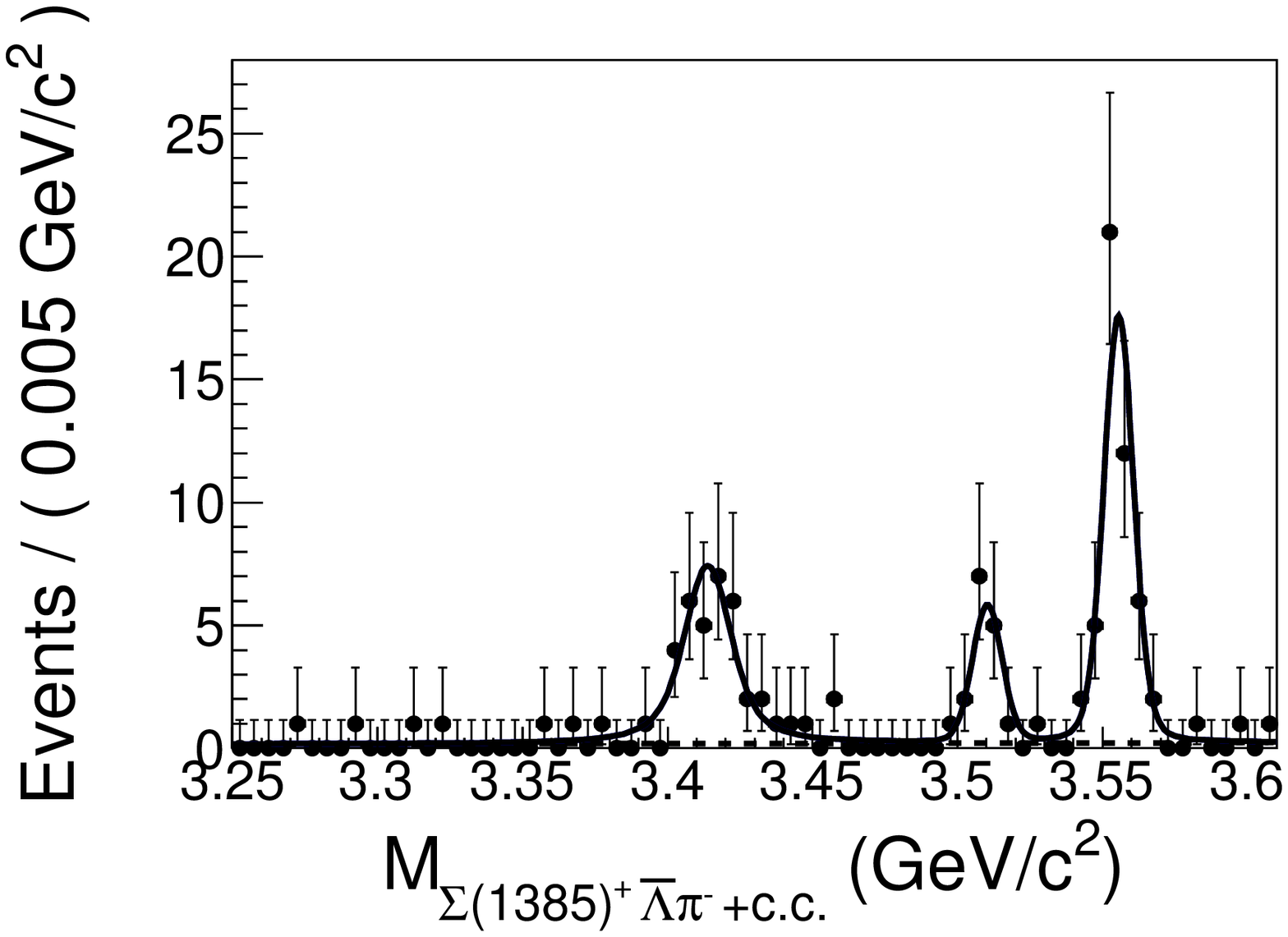}
\put(25, 40){(b)}
\end{overpic}
\end{minipage}
\begin{minipage}[t]{0.45\linewidth}
\begin{overpic}[height=0.20\textheight,width=1.0\linewidth]
{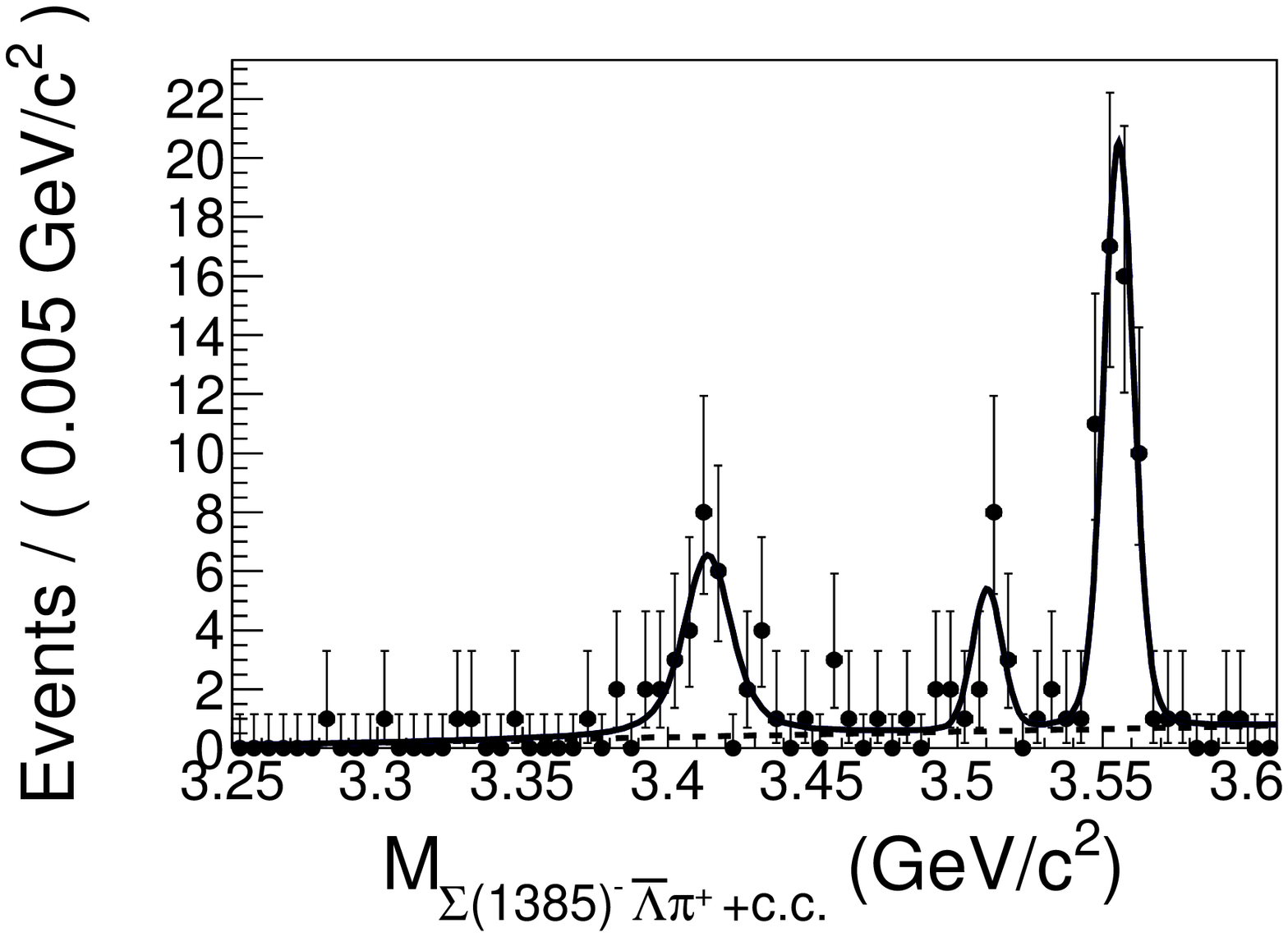}
\put(25, 40){(c)}
\end{overpic}
\end{minipage}
\begin{minipage}[t]{0.45\linewidth}
\begin{overpic}[height=0.20\textheight,width=1.0\linewidth]
{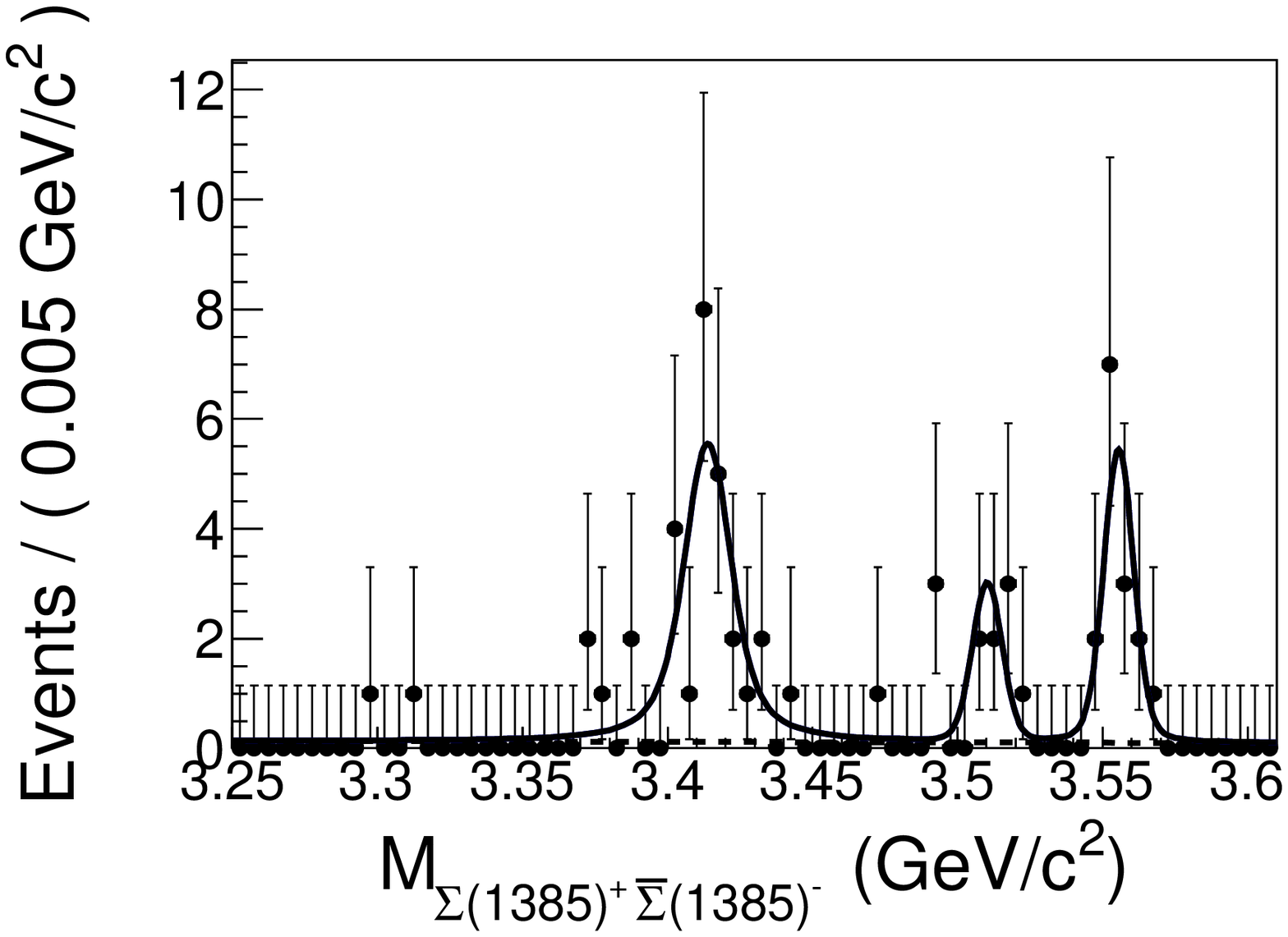}
\put(25, 40){(d)}
\end{overpic}
\end{minipage}
\begin{minipage}[t]{0.45\linewidth}
\begin{overpic}[height=0.20\textheight,width=1.0\linewidth]
{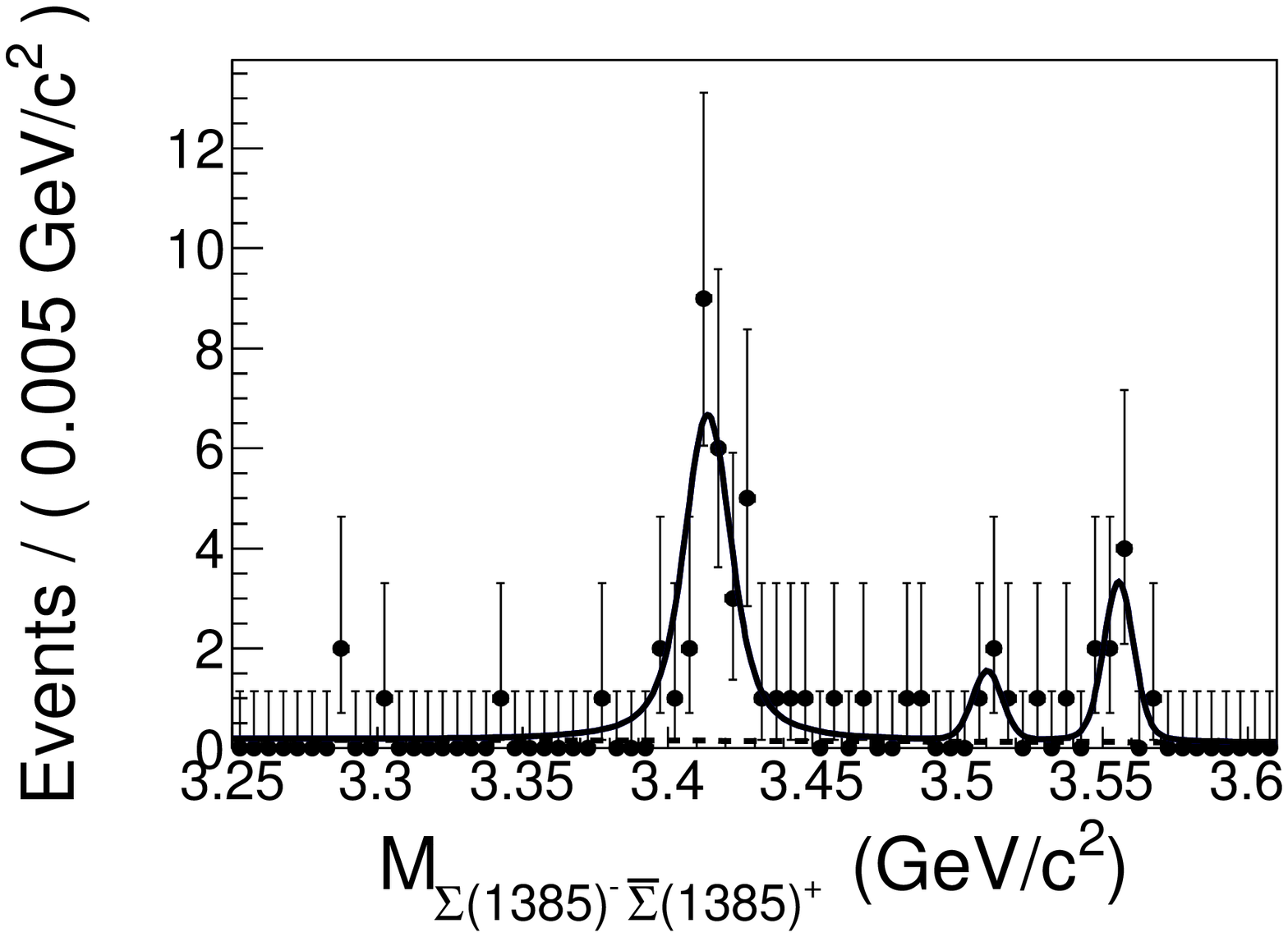}
\put(25, 40){(e)}
\end{overpic}
\end{minipage}
\begin{minipage}[t]{0.45\linewidth}
\begin{overpic}[height=0.20\textheight,width=1.0\linewidth]
{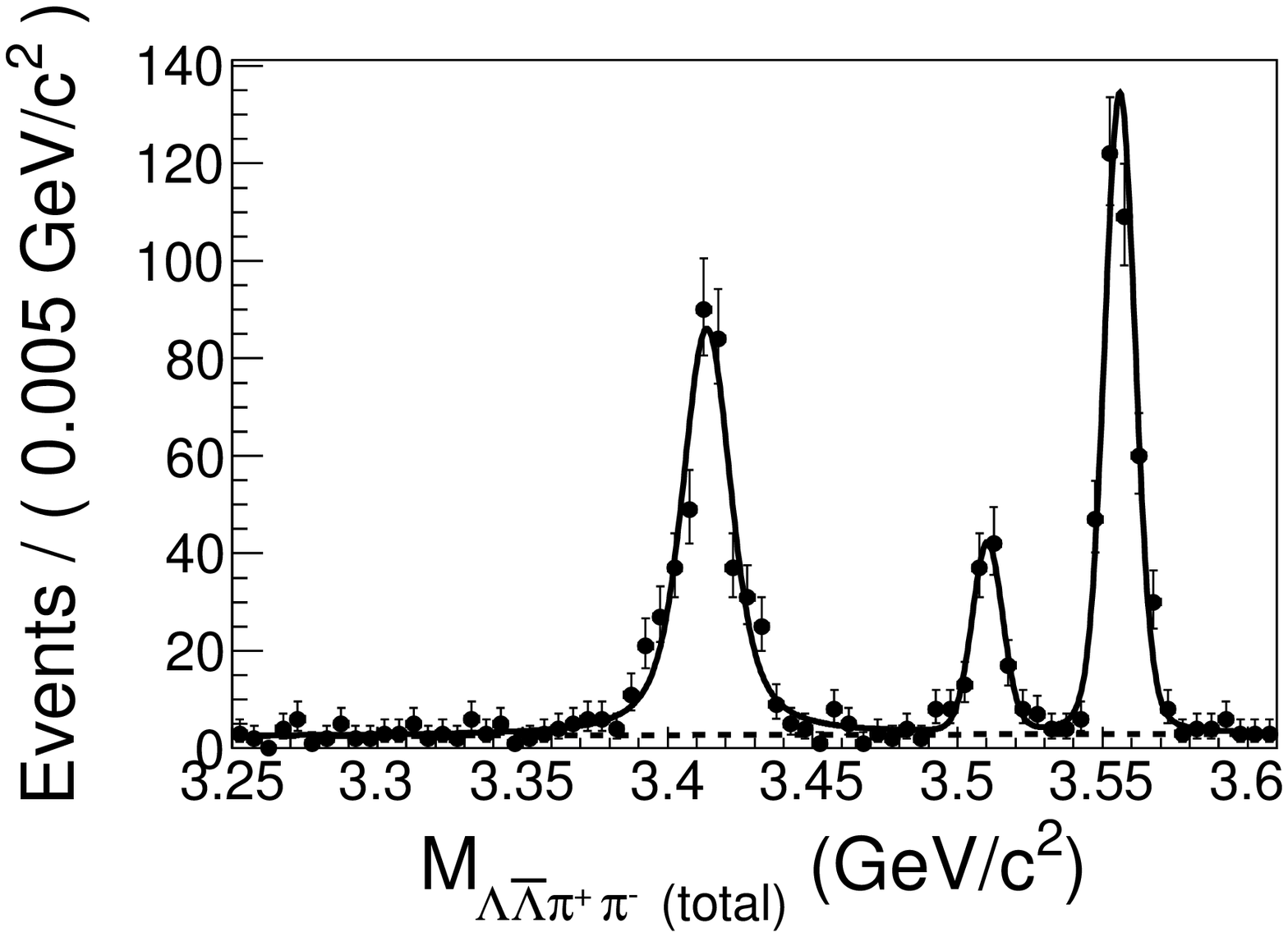}
\put(25, 40){(f)}
\end{overpic}
\end{minipage}
\renewcommand{\figurename}{FIG.}
\caption{The invariant mass distributions  of $\Lambda\bar{\Lambda}\pi^{+}\pi^{-}$
  in the following data samples: (a) data set-1, (b) data set-2, (c) data set-3, %
    (d) data set-4, (e) data set-5 and (f) total data set. The selections of
    data set-$j$ ($j=1,\cdots5$) are defined in Sec.~\ref{data_analysis}.
    Points with error bars are data. The solid curves show the sum of the fitted
    curves, while the dashed lines are the backgrounds.
}
{\label{figure_fitplot_2L2Pi_all}}
\end{figure}

Due to the broad width and the long tails of the $\Sigma(1385)$, the
$\Sigma(1385)$ and non-$\Sigma(1385)$ events feed into the
non-$\Sigma(1385)$ and $\Sigma(1385)$ regions. As a result, the
$\chi_{cJ}$ events that decay into the above five processes cannot be
completely separated using invariant mass regions alone.  In this
study, we separate the data into five independent categories, with
data set labels set-$j$ ($j=1, \cdots,\ 5$) defined as follows:
\begin{enumerate}[(i)]
\item Data set-1:
the category to detect the non-resonant process 1. That is, events with
$M_{\Lambda\pi^{+}}$, $M_{\bar{\Lambda}\pi^{-}}$, $M_{\Lambda\pi^{-}}$ and
$M_{\bar{\Lambda}\pi^{+}}$  all in non-$\Sigma(1385)$ regions. The invariant
mass spectrum of $\Lambda\bar{\Lambda}\pi^{+}\pi^{-}$ is displayed in
Fig. \ref{figure_fitplot_2L2Pi_all} (a);

\item Data set-2: the category to detect the single resonant
$\Sigma(1385)^+(\bar{\Sigma}(1385)^-)$ process 2.  That is, events
with $M_{\Lambda\pi^{+}}/M_{\bar{\Lambda}\pi^{-}}$ in the
$\Sigma(1385)$ signal region and with
$M_{\bar{\Lambda}\pi^{-}}$($M_{\Lambda\pi^{+}}$), %
$M_{\Lambda\pi^{-}}$, $M_{\bar{\Lambda}\pi^{+}}$ in
non-$\Sigma(1385)$ regions are required.  The two types of events in
this category are combined and displayed in
Fig. \ref{figure_fitplot_2L2Pi_all} (b);

\item Data set-3: the category to detect the single resonant
$\Sigma(1385)^-(\bar{\Sigma}(1385)^+)$ process 3.  Similarly, events
with $M_{\Lambda\pi^{-}}/M_{\bar{\Lambda}\pi^{+}}$ in $\Sigma(1385)$
signal region and with
$M_{\bar{\Lambda}\pi^{+}}$($M_{\Lambda\pi^{-}}$), %
$M_{\Lambda\pi^{+}}$, $M_{\bar{\Lambda}\pi^{-}}$ in
non-$\Sigma(1385)$ regions are required.  The two types of events in
this category are combined and displayed in
Fig. \ref{figure_fitplot_2L2Pi_all} (c);

\item  Data set-4: the category to detect process 4.  Events  with
$M_{\Lambda\pi^{+}}$, $M_{\bar{\Lambda}\pi^{-}}$ in  $\Sigma(1385)$
signal region and $M_{\Lambda\pi^{-}}$, $M_{\bar{\Lambda}\pi^{+}}$
in non-$\Sigma(1385)$ region are selected and displayed in
Fig. \ref{figure_fitplot_2L2Pi_all} (d);

\item  Data set-5: the category to detect  process 5. Events  with
$M_{\Lambda\pi^{-}}$, $M_{\bar{\Lambda}\pi^{+}}$ in  $\Sigma(1385)$
signal region and with $M_{\Lambda\pi^{+}}$, $M_{\bar{\Lambda}\pi^{-}}$
in  non-$\Sigma(1385)$ region are selected and displayed in
Fig. \ref{figure_fitplot_2L2Pi_all} (e).
\end{enumerate}
The yield in each data set is estimated
by a fit to the $\chi_{cJ}$ peaks, and the yields of each process
in the full phase space will be disentangled with
Eq.~(\ref{equation_2L2Pi_tot}), as described in
Sec.~\ref{section_branching_fractions}.

The $\chi_{cJ}$ signal events are clearly observed in each category, as
shown in Fig. \ref{figure_fitplot_2L2Pi_all} (a)--(e).  In the fits
of the $\chi_{cJ}$ in each data set category, a Breit-Wigner function
convolved with a Gaussian resolution function is used to describe
$\chi_{cJ}$ peaks, while a 1st-order polynomial line is used to model
the background distribution.  The $\chi_{cJ}$  invariant mass
parameters are allowed to float, while the $\chi_{cJ}$ widths are
fixed to the PDG values~\cite{pdg12}. The Gaussian parameters are obtained
from MC simulation of detector responses. A simultaneous unbinned maximum
likelihood method is applied, and the fit results are listed in
Table \ref{table_fit_results}.

\begin{table}[t!]
\renewcommand{\tablename}{TABLE}
\begin{center}
  \caption{
    The number of fitted $\chi_{cJ}$ events in each data set-$j$ ($j=1, \cdots,\ 5$)
      and the total data set.  $n_{j}$ is the number of fitted $\chi_{cJ}$ events in
      data set-${j}$.  $n_{\mathrm{tot}}$ is that in the total data sample.
  }
\label{table_fit_results}
\input{./table_fit_results.tex}
\end{center}
\end{table}


\subsection{Background Study}

A 106 million inclusive $\psi^\prime$-decay MC sample is used to
investigate possible $\psi^\prime$ decay backgrounds. No peaking
backgrounds are observed, as shown in
Fig.~\ref{figure_signal_lam_chicj}(d). Since a large proportion of the
$\chi_{cJ}$ decays are poorly known and their simulations based on the
BESIII LundCharm model have large uncertainty, we investigate possible
underestimated peaking backgrounds beneath the $\chi_{cJ}$ peaks. One
major source could be from $\chi_{cJ}\rightarrow\Lambda K^{*+}
\bar{p}\rightarrow p \bar{p}\pi^{+}\pi^{-} K^{0}_{s}\rightarrow p
\bar{p}2\pi^{+}2\pi^{-} \ (c.c.)$; however, the $\pi^{+}$ and
$\pi^{-}$ invariant mass distributions of candidate events were
examined, and no evidence of a $K_{s}$ peak was found.  Therefore,
  negligible peaking background is assumed in this study.  A study of
  the continuum data did not reveal any non-$\psi^\prime$ decay
  backgrounds.

  \subsection{Calculation of  Branching Ratios}\label{section_branching_fractions}

  To calculate the branching ratios of each mode in
  $\chi_{cJ}\to\Lambda\bar{\Lambda}\pi^{+}\pi^{-}$ decay, one has to compute the
  efficiency-corrected number of $\chi_{cJ}$ decays.  The numbers of
  $\chi_{cJ}$ events in data set-$j$, which is selected to detect
  process $i$,  also consists of events from the other processes. We
  describe the number of  events of process $i$ in  data set-$j$ as

  \begin{equation}\label{equation_2L2Pi_tot}
  \left(
      \begin{array}{cc c c c}
      \varepsilon_{11} & \varepsilon_{12} & \varepsilon_{13} & \varepsilon_{14} & \varepsilon_{15} \\
      \varepsilon_{21} & \varepsilon_{22} & \varepsilon_{23} & \varepsilon_{24} & \varepsilon_{25} \\
      \varepsilon_{31} & \varepsilon_{32} & \varepsilon_{33} & \varepsilon_{34} & \varepsilon_{35} \\
      \varepsilon_{41} & \varepsilon_{42} & \varepsilon_{43} & \varepsilon_{44} & \varepsilon_{45} \\
      \varepsilon_{51} & \varepsilon_{52} & \varepsilon_{53} & \varepsilon_{54} & \varepsilon_{55}
      \end{array}
      \right)
  \left(
      \begin{array}{c}
      N_{1} \\ N_{2} \\ N_{3} \\ N_{4} \\ N_{5}
      \end{array}
      \right)
  =
  \left(
      \begin{array}{c}
      n_{1} \\ n_{2} \\ n_{3} \\ n_{4} \\ n_{5}
      \end{array}
      \right),
  \end{equation}
  \noindent where $N_{i}$ is the efficiency-corrected number of events
  of process $i$, $n_{j}$ are the numbers of $\chi_{cJ}$ events in the
  data set-$j$ (as listed in Table \ref{table_fit_results}), and
  $\varepsilon_{ji}$ denotes the efficiency of process $i$ being
  selected in data set-$j$, obtained with MC simulation.  In practice,
  the $\chi_{cJ}$ signals are fitted in the five data sets
  simultaneously, and with the constraint of the three efficiency
  matrices, $N_{1}$ -- $N_{5}$ are obtained by the fit. Equations
  \eqref{formula_br_2L2Pi}--\eqref{formula_br_SS} are used to
  calculate branching ratios ($\mathcal{B}$) of the signal processes,
  and the results are listed in Table~\ref{table_final_results}.
  The significance of each decay mode, which is estimated using
  Eq. (\ref{equation_significance}), is listed in
  Table~\ref{table_final_results}.   Here $L_{m}$ is the likelihood
  of the simultaneous fit, while $L_{m(N_{j}=0)}$ is the likelihood of
  the fit with the assumption that $N_{j}$ is equal to zero.

  \begin{equation}\label{formula_br_2L2Pi}
  \mathcal{B}{(\chi_{cJ}\rightarrow\Lambda \bar{\Lambda}\pi^{+}\pi^{-}
      (\mathrm{non-resonant}))} =
  \frac{N_{1}}
{
  N_{_{\psi^\prime}}
  \mathcal{B}{(\psi^\prime\rightarrow\gamma\chi_{cJ})}
  \mathcal{B}{(\Lambda\rightarrow p \pi)^{2}}
}
\end{equation}
\begin{equation}\label{formula_br_SLPi}
\begin{split}
\begin{aligned}
\mathcal{B}{(\chi_{cJ}\rightarrow\Sigma(1385)^{+(-)}\bar{\Lambda}\pi^{-(+)}+c.c. )} =
\frac{N_{2(3)}}
{
  N_{_{\psi^\prime}}
  \mathcal{B}{(\psi^\prime\rightarrow\gamma\chi_{cJ})}
  \mathcal{B}{(\Sigma(1385)\rightarrow\Lambda\pi)}
} \\
    \cdot \frac{1.0}
{
  \mathcal{B}{(\Lambda\rightarrow p \pi)^{2}}
}
\end{aligned}
\end{split}
\end{equation}
\begin{equation}\label{formula_br_SS}
\begin{split}
\begin{aligned}
\mathcal{B}{(\chi_{cJ}\rightarrow\Sigma(1385)^{+(-)}\bar{\Sigma}(1385)^{-(+)})} =
\frac{N_{4(5)}}
{
  N_{_{\psi^\prime}}
  \mathcal{B}{(\psi^\prime\rightarrow\gamma\chi_{cJ})}
  \mathcal{B}{(\Sigma(1385)\rightarrow\Lambda\pi)^{2}}
} \\
    \cdot \frac{1.0}
{
  \mathcal{B}{(\Lambda\rightarrow p \pi)^{2}}
}
\end{aligned}
\end{split}
\end{equation}
\begin{equation}\label{equation_significance}
S_{j} = \sqrt{2\times(\mathrm{ln}L_{m} - \mathrm{ln}L_{m(N_{j}=0)})}
\end{equation}

\begin{figure}[t!]
  \begin{minipage}[t]{0.32\linewidth}
  \begin{overpic} [height=0.20\textheight,width=1.0\linewidth]
{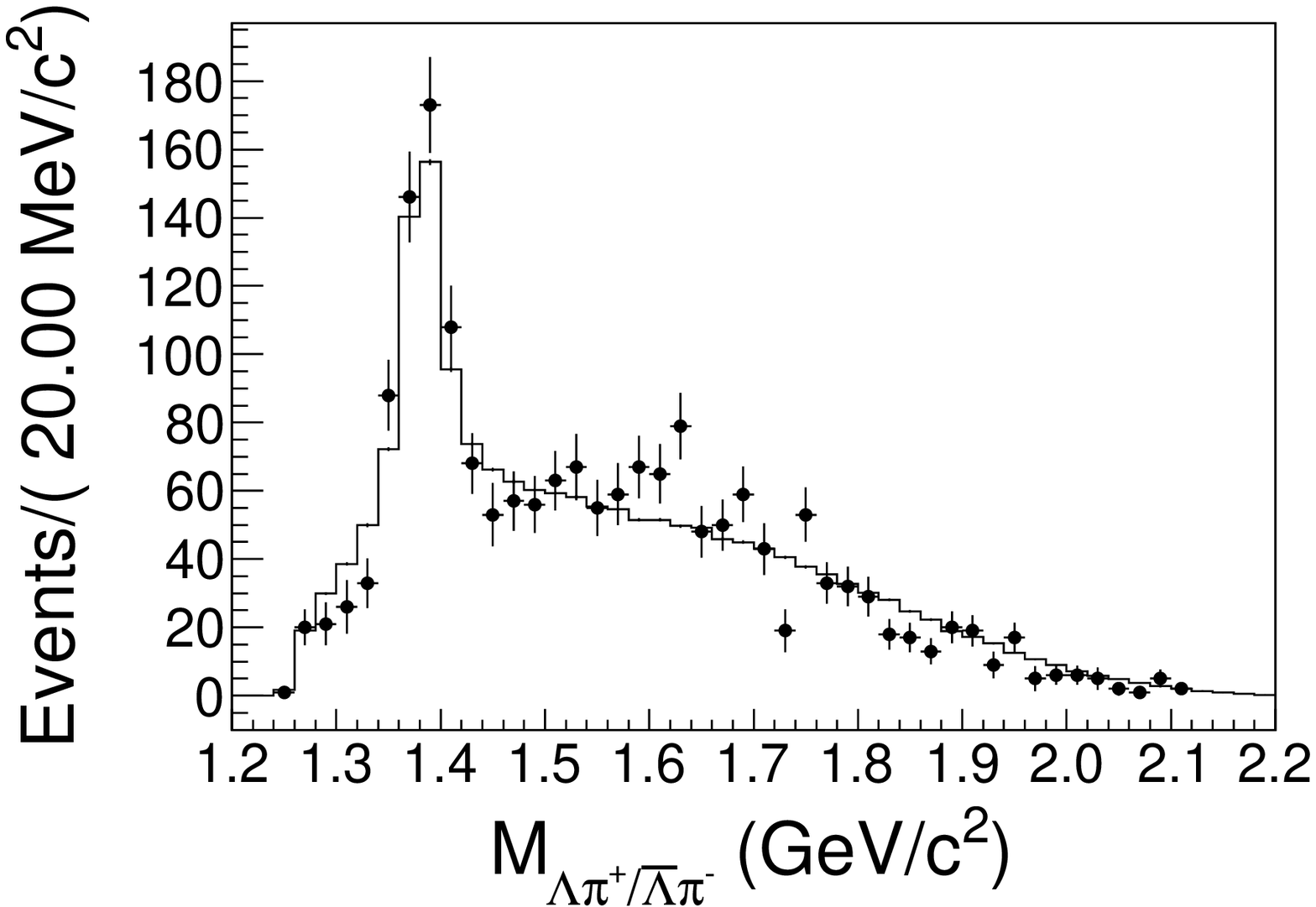}
\put(70, 60){(a)}
\end{overpic}
\end{minipage}
\begin{minipage}[t]{0.32\linewidth}
\begin{overpic} [height=0.20\textheight,width=1.0\linewidth]
{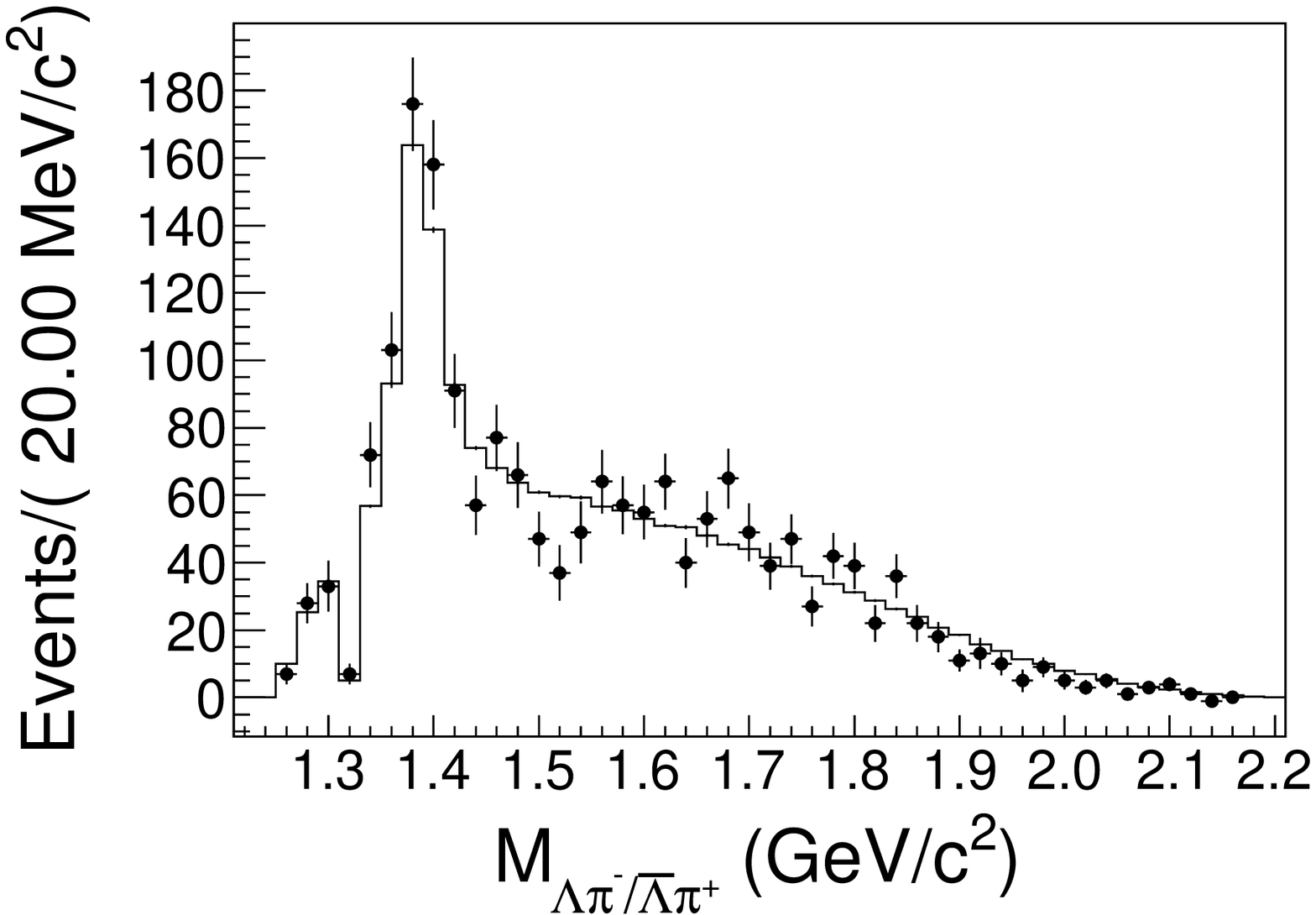}
\put(70, 60){(b)}
\end{overpic}
\end{minipage}
\begin{minipage}[t]{0.32\linewidth}
\begin{overpic} [height=0.20\textheight,width=1.0\linewidth]
{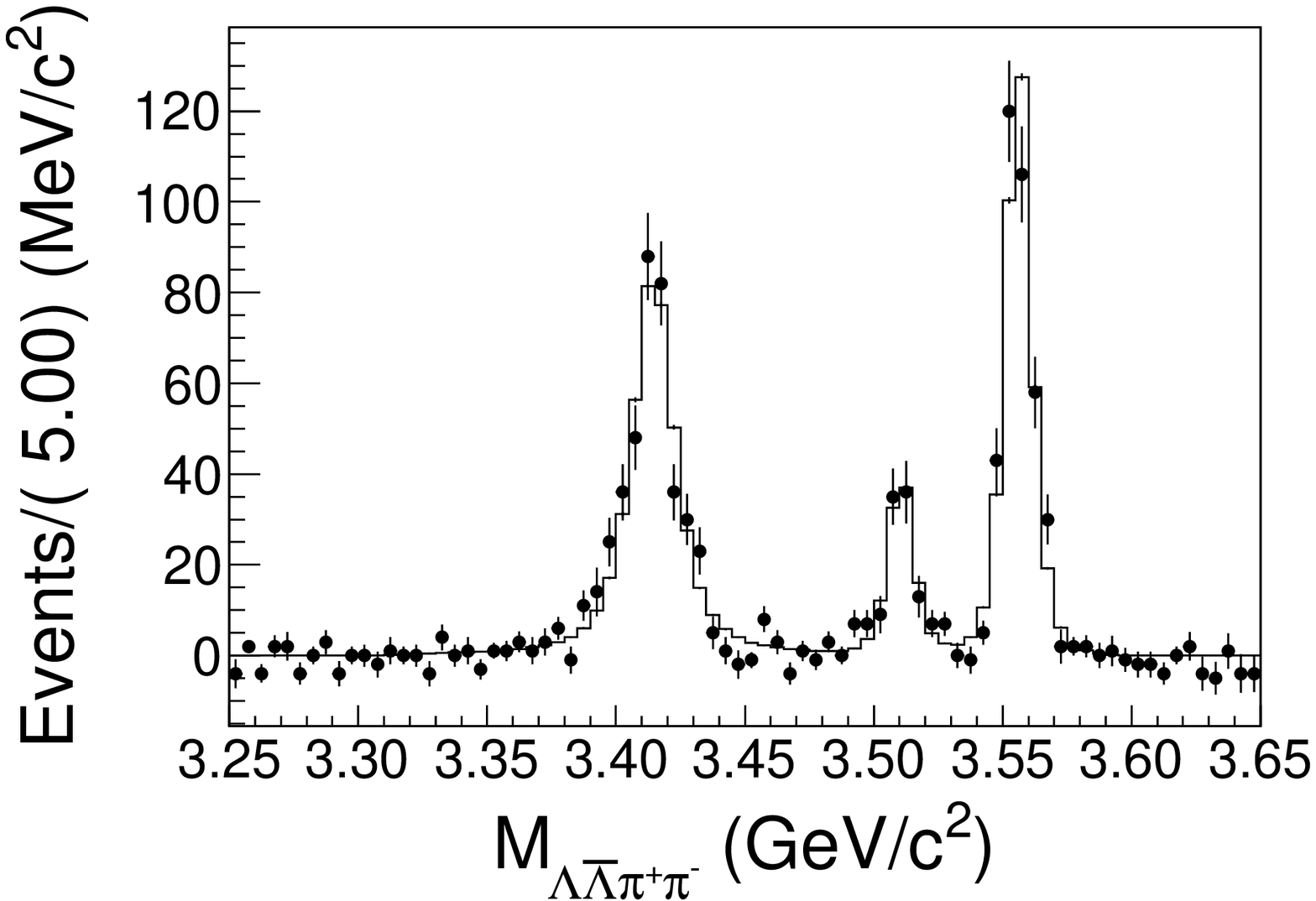}
\put(30, 60){(c)}
\end{overpic}
\end{minipage}
\renewcommand{\figurename}{FIG.}
\caption{
  The invariant mass distributions of (a) $\Lambda\pi^{+}/\bar{\Lambda}\pi^{-}$,
    (b) $\Lambda\pi^{-}/\bar{\Lambda}\pi^{+}$ and (c) $\Lambda\bar{\Lambda}\pi^{+}\pi^{-}$.
    Points with error bars are the data with subtraction of the backgrounds,
    while solid lines are the MC simulation of the signals;
    The backgrounds subtracted are estimated from inclusive MC.
    The signal components are scaled based on their branching ratios measured in this work.
    The data within $1.312$\,GeV/$c^{2}<M_{\Lambda\pi^{-}(\bar{\Lambda}\pi^{+})}<1.331$\,GeV/$c^{2}$ are removed to reject the $\Xi^{\pm}$ candidates.
}
{\label{figure_cp_signal_s1385}}
\end{figure}

As shown in Fig.~\ref{figure_cp_signal_s1385}, the sum of measured components in the decays of $\chi_{cJ}$ into the final states $\Lambda\bar{\Lambda}\pi^+\pi^-$ in MC simulation agrees well with the data. This supports the credibility of the decomposition into the different components described above.

\subsection{$\chi_{cJ}\rightarrow\Lambda\bar{\Lambda}\pi^{+}\pi^{-}$ (total)}\label{section_2L2Pi_tot}

Based on the selection criteria in Sec. \ref{event_selection}, the
process $\chi_{cJ}\rightarrow\Lambda\bar{\Lambda}\pi^{+}\pi^{-}$
(total), including the intermediate-resonant processes, is studied.
The $\Lambda\bar{\Lambda}\pi^{+}\pi^{-}$ invariant mass distributions
and the fit are displayed in Fig. \ref{figure_fitplot_2L2Pi_all} (f),
    while the fit results are listed in Table \ref{table_fit_results}.
    According to the measured branching ratios of the intermediate
    resonances in this analysis, signal MC samples are generated.  This
    makes the momentum distributions of the final particles in the MC
    sample similar to those in experimental data and allows the
    determination of the overall detection efficiency,
    $\varepsilon_{\mathrm{tot}}$, of the sum of all the processes with the
    same final states $\psi^{\prime}\rightarrow\gamma\chi_{cJ}
    \rightarrow\gamma\Lambda\bar{\Lambda}\pi^{+}\pi^{-}$.  The branching
    ratio of $\chi_{cJ}\rightarrow\Lambda\bar{\Lambda}\pi^{+}\pi^{-}$
    (total) is calculated with the formula

    \begin{equation}\label{formula_br_2L2Pi_tot}
    \mathcal{B}{(\chi_{cJ}\rightarrow\Lambda \bar{\Lambda}\pi^{+}\pi^{-}(\mathrm{total)})} =
    \frac{n_{\mathrm{tot}}}
{
  \varepsilon_{\mathrm{tot}}
  \cdot N_{_{\psi^\prime}}
  \mathcal{B}{(\psi^\prime\rightarrow\gamma\chi_{cJ})}
  \mathcal{B}{(\Lambda\rightarrow p \pi)^{2}}
}
\end{equation}

    \begin{table}
    \renewcommand{\tablename}{TABLE}
    \begin{center}
    \caption{
      Results of the branching ratios ($\times10^{-5}$) for different decay modes.
	`UL' stands for the upper limit of the branching ratio at the $90\%$ C.L.
	`$S$' stands for the statistical significance.
	The first errors are statistical and the second systematic.
    }\label{table_final_results}
{\footnotesize
  \input{./table_final_results.tex}
}
\end{center}
\end{table}

\section{Systematic Uncertainty}

The systematic uncertainties in this analysis are summarized in Table
\ref{table_sys_err_source}.  Sources of systematic uncertainty include
$\Lambda/\bar{\Lambda}$ reconstruction, $\pi^{\pm}$ tracking, photon
detection, four-momentum constraint kinematic fitting, background
rejection, $\chi_{cJ}$ signal fitting, the number of $\psi^{\prime}$
events and branching ratios cited from the PDG~\cite{pdg12}. Charged
$\pi$ tracking and photon detection systematic errors are studied
following the methods in Refs.~\cite{BAM0003, bes3}.

\begin{table}[t!]
\renewcommand{\tablename}{TABLE}
\begin{center}
\caption{Sources of systematic  uncertainties.}\label{table_sys_err_source}
\input{./table_sys_err_source.tex}
\end{center}
\end{table}

For the systematic uncertainty due to $\Lambda/\bar{\Lambda}$
reconstruction, J$/\psi\rightarrow\Lambda\bar{\Lambda}\pi^{+}\pi^{-}$
and
$\psi^\prime\rightarrow\pi^{+}\pi^{-}$J/$\psi\rightarrow\Lambda\bar{\Lambda}\pi^{+}\pi^{-}$
are used to select a $\Lambda/\bar{\Lambda}$ control sample.
$\Lambda/\bar{\Lambda}$ reconstruction efficiency is calculated by
taking the ratio of the fitted $\Lambda/\bar{\Lambda}$ yields in
the missing mass spectrum before and after $\Lambda/\bar{\Lambda}$ is
found.  $\Lambda/\bar{\Lambda}$ reconstruction efficiencies consist of
tracking efficiency of the daughter particles and the
vertex-constraint of $\Lambda/\bar{\Lambda}$.  The differences in the
efficiencies between experimental data and the MC sample are included
in the systematic uncertainties.

To study the efficiency of the kinematic fitting in the four-momentum
constraint, event candidates for the three processes $\psi^\prime
\rightarrow\gamma\Lambda\bar{\Lambda}\pi^{+}\pi^{-}$, $\psi^\prime
\rightarrow\gamma\chi_{cJ}\rightarrow \gamma3(\pi^{+}\pi^{-})$
and $\psi^\prime\rightarrow J/\psi\pi^{+}\pi^{-}\rightarrow
3(\pi^{+}\pi^{-})\pi^{0} \rightarrow 3(\pi^{+}\pi^{-})2\gamma$
are used as control samples.  The ratio of the event rates before
and after the kinematic fitting is taken as the efficiency of the
kinematic fitting.  These efficiencies are calculated both in
experimental data and in the MC sample, and their difference determines
the uncertainty of the kinematic fitting.

For the rejection of the resonances J/$\psi$,
    $\Sigma^{0}$/$\bar{\Sigma}^{0}$ and $\Xi^{\pm}$, different J/$\psi$,
    $\Sigma^{0}$, $\bar{\Sigma}^{0}$ and $\Xi^{\pm}$ mass region
    requirements are applied ranging from 3$\sigma$, 3.5$\sigma$ to
    4$\sigma$, where $\sigma$ is the detector resolution. The largest
    deviation on the branching ratios is taken as the systematic
    uncertainty.  The systematic uncertainty of the fitting method is
    obtained by changing the fitting range, the shape of the backgrounds,
    and changing the detector resolution from the value obtained with MC
    simulation to that obtained by fitting with a free parameter.  The
    relative uncertainty of the estimated number of $\psi^\prime$ is
    $4.0\%$~\cite{BAM0003}. The uncertainty of the branching ratios of
    intermediate decays are taken from the PDG~\cite{pdg12}.  The total
    systematic uncertainty is obtained by summing all the individual
    uncertainties in quadrature.

\section{Results and Discussion}
The branching ratios of $\chi_{cJ}$ decays to $\Sigma(1385)^{\pm}
\bar{\Sigma}(1385)^{\mp}$, $\Sigma(1385)^{\pm}\bar{\Lambda}\pi^{\mp} +
c.c.$ and $\Lambda\bar{\Lambda}\pi^{+}\pi^{-}$ (with or without the
    $\Sigma(1385)$ resonance) are measured with $106$ million
$\psi^\prime$ decay events collected at BESIII. The results are listed
in Table \ref{table_final_results}.  The process
$\chi_{cJ}\rightarrow\Lambda\bar{\Lambda}\pi^{+}\pi^{-}$ is observed
for the first time. Evidence of
$\chi_{c0}\rightarrow\Sigma(1385)^{\pm}\bar{\Sigma}(1385)^{\mp}$,
  which strongly violates the helicity selection rule, is presented.
  The branching ratios of
  $\chi_{c1,2}\rightarrow\Sigma(1385)^{\pm}\bar{\Sigma}(1385)^{\mp}$ are
  consistent with the theoretical predictions~\cite{PRAv674p185}.

  \section{Acknowledgments}
  The BESIII collaboration thanks the staff of BEPCII and the computing
  center for their hard work. This work is supported in part by the
  Ministry of Science and Technology of China under Contract No.
  2009CB825200; National Natural Science Foundation of China (NSFC)
  under Contracts Nos.
  10625524, 10821063, 10825524, 10835001, 10935007, 10905091, 11079030, %
  11125525; Joint Funds of the National Natural
  Science Foundation of China under Contracts Nos. 11079008, 11179007;
  the Chinese Academy of Sciences (CAS) Large-Scale Scientific Facility
  Program; CAS under Contracts Nos. KJCX2-YW-N29, KJCX2-YW-N45; 100
  Talents Program of CAS;
  Research Fund for the Doctoral Program of Higher Education of China
  under Contract No. 20093402120022;
  Istituto Nazionale di Fisica Nucleare, Italy;
  U. S. Department of Energy under Contracts Nos. DE-FG02-04ER41291, %
  DE-FG02-91ER40682, DE-FG02-94ER40823; U.S. National Science
  Foundation; University of Groningen (RuG) and the Helmholtzzentrum
  fuer Schwerionenforschung GmbH (GSI), Darmstadt; WCU Program of
  National Research Foundation of Korea under Contract No.
  R32-2008-000-10155-0.

  \newpage

  \end{document}

%% file: table_fit_results.tex
\begin{tabular}{c| c| C{6em} C{6em} C{6em}}
\hline
\hline
\multicolumn{2}{c|}{Number of events} & \multicolumn{1}{c}{$\chi_{c0}$} & \multicolumn{1}{c}{$\chi_{c1}$} & \multicolumn{1}{c}{$\chi_{c2}$}   \\
\hline
\multicolumn{2}{c|}{$n_{1}$}  &10.8$\pm$3.8&12.7$\pm$3.9&36.4$\pm$6.4\\
\multicolumn{2}{c|}{$n_{2}$}  &36.4$\pm$6.7&14.7$\pm$4.1&47.6$\pm$7.2\\
\multicolumn{2}{c|}{$n_{3}$}  &30.9$\pm$6.6&12.5$\pm$4.1&54.4$\pm$7.9\\
\multicolumn{2}{c|}{$n_{4}$}  &27.4$\pm$5.9&7.6$\pm$3.2&14.6$\pm$4.0\\
\multicolumn{2}{c|}{$n_{5}$}  &32.8$\pm$6.3&3.6$\pm$2.2&8.7$\pm$3.3\\
\multicolumn{2}{c|}{ $n_{\mathrm{tot}}$}
 &426$\pm$23&105$\pm$11&371$\pm$20\\
\hline
\hline
\end{tabular}

%% file: table_final_results.tex
\begin{tabular}{c| c| c| c| c| c| c| c| c| c}
\hline
\hline
\multirow{2}{*}{ $\chi_{cJ}$ decay mode} &
\multicolumn{3}{c|}{$\chi_{c0}$}&
\multicolumn{3}{c|}{$\chi_{c1}$}&
\multicolumn{3}{c}{$\chi_{c2}$}
\\
\cline{2-10}
\multicolumn{1}{c|}{} &
\multicolumn{1}{c|}{$\mathcal{B}$}&
\multicolumn{1}{c|}{UL}&
\multicolumn{1}{c|}{$S$}&
\multicolumn{1}{c|}{$\mathcal{B}$}&
\multicolumn{1}{c|}{UL}&
\multicolumn{1}{c|}{$S$}&
\multicolumn{1}{c|}{$\mathcal{B}$}&
\multicolumn{1}{c|}{UL} &
\multicolumn{1}{c}{$S$}
\\
\hline
$\Lambda\bar{\Lambda}\pi^{+}\pi^{-}$ (w/o $\Sigma(1385)$) & 28.6$\pm$12.6$\pm$2.7 &  $<$54 & 2.2 &  26.2$\pm$5.5$\pm$3.3 &   &  4.8 &  71.8$\pm$14.5$\pm$8.2 &   &  6.4 \\
$\Sigma(1385)^{+}\bar{\Lambda}\pi^{-}+c.c.$ & 34.8$\pm$13.2$\pm$3.4 &  $<$55 & 2.2 &   &  $<$14 & 0.3 &  23.6$\pm$11.8$\pm$2.7 &  $<$42 & 1.7 \\
$\Sigma(1385)^{-}\bar{\Lambda}\pi^{+}+c.c.$ & 24.6$\pm$12.7$\pm$2.4 &  $<$50 & 1.6 &   &  $<$14 & 0.0 &  37.8$\pm$11.8$\pm$4.4 &  $<$61 & 2.6 \\
$\Sigma(1385)^{+}\bar{\Sigma}(1385)^{-}$ & 16.4$\pm$5.7$\pm$1.6 &   &  3.1 &  4.4$\pm$2.5$\pm$0.6 &  $<$10 & 1.9 &  7.9$\pm$4.0$\pm$0.9 &  $<$17 & 2.0 \\
$\Sigma(1385)^{-}\bar{\Sigma}(1385)^{+}$ & 23.5$\pm$6.2$\pm$2.3 &   &  4.3 &   &  $<$5.7 & 0.9 &   &  $<$8.5 & 0.0 \\
$\Lambda\bar{\Lambda}\pi^{+}\pi^{-}$(total) & 119.0$\pm$6.4$\pm$11.4 &  &  $>10$  &  31.1$\pm$3.4$\pm$3.9 &  &  $>10$  &  137.0$\pm$7.6$\pm$15.7 &  &  $>10$  \\
\hline
\hline
\end{tabular}

%% file: table_sys_err_source.tex
\begin{tabular}{ c| C{6em} C{6em} C{6em}}
\hline
\hline
\multirow{2}{*}{Sources} & 
\multicolumn{3}{c}{Relative systematic uncertainty (\%)}  
\\
\cline{2-4}
\multicolumn{1}{c|}{ } & 
\multicolumn{1}{c}{$\chi_{c0}$} &   
\multicolumn{1}{c}{$\chi_{c1}$} &   
\multicolumn{1}{c}{$\chi_{c2}$}    
\\
\hline
\multicolumn{1}{c|}{$\Lambda\bar{\Lambda}$ reconstruction}  & 3.5 & 3.5 & 3.5 \\
\multicolumn{1}{c|}{$\pi^{\pm}$ tracking (not from $\Lambda\bar{\Lambda}$)}  & 2.0 & 2.0  &2.0 \\
\multicolumn{1}{c|}{photon detection}  & 1.0 & 1.0  &1.0 \\
\multicolumn{1}{c|}{kinematic fitting}  & 1.0 & 1.0  &1.0 \\
\multicolumn{1}{c|}{vetoing background}  & 4.9 & 2.6  &4.4 \\
\multicolumn{1}{c|}{fitting method}  & 4.9 & 10.1  &7.9 \\
\multicolumn{1}{c|}{the number of $\psi^\prime$}  & 4.0 & 4.0 & 4.0 \\
 \multicolumn{1}{c|}{$\mathcal{B}$($\psi^\prime\rightarrow\gamma\chi_{cJ} $) }& 3.2 & 4.3 & 4.0 \\
 \multicolumn{1}{c|}{$\mathcal{B}$($\Sigma(1385)^{\pm}\rightarrow\Lambda\pi^{\pm}$)}  & 1.7 & 1.7 & 1.7 \\
\multicolumn{1}{c|}{$\mathcal{B}$($\bar{\Sigma}(1385)^{\mp}\rightarrow\bar{\Lambda}\pi^{\mp}$)}  & 1.7 & 1.7 & 1.7 \\
\multicolumn{1}{c|}{total}  & 9.6 & 12.7  &11.4 \\
\hline
\hline
\end{tabular}